\documentclass[a4paper,onecolumn,accepted=2026-01-17]{quantumarticle}
\pdfoutput=1
\usepackage{orcidlink} 

\usepackage[utf8]{inputenc}
\usepackage[toc,page]{appendix}
\usepackage{amsfonts}
\usepackage{color}
\usepackage{amsmath}
\usepackage{graphicx}
\usepackage{amssymb}
\usepackage{hyperref}
\usepackage{color}
\usepackage{mathrsfs}
\usepackage{isomath}
\usepackage{amsthm}
\usepackage{epstopdf}
\usepackage{txfonts}
\usepackage{dsfont}
\usepackage{ulem}
\usepackage{physics} 
\usepackage{tikz} 
\allowdisplaybreaks[4]

\usepackage{bbold}
\usepackage{mathtools}

\usepackage[numbers,sort&compress]{natbib}

\newcommand{\ignore}[1]{} 

\newcommand{\hil}{\mathcal{H}}
\newcommand{\obsset}[1]{\boldsymbol{#1}}

\renewcommand{\tr}{\mathrm{tr}} 

\newcommand{\be}{\begin{equation}}
\newcommand{\ee}{\end{equation}}
\newcommand{\eea}{\end{eqnarray}}
\newcommand{\bea}{\begin{eqnarray}}

\newcommand{\av}[1]{\ensuremath{\langle{#1} \rangle}}

\newcommand{\Cov}{{\rm Cov}}
\renewcommand{\vec}[1]{\boldsymbol{#1}}

\newcommand{\id}{\mathbb{1}}

\newcommand{\Covn}[1]{{\rm Cov}_{\varrho_{#1}}(\obsset{g}_{#1})}

\newcommand{\C}{X_\varrho}

\newcommand{\Cp}{X_\psi}

\newcommand{\DiffMat}{\mathfrak \Delta}

\newcommand{\SN}{\mathcal{SN}}

\newcommand{\SNV}{\boldsymbol{\mathcal{SN}}}

\newcommand{\Vfeas}{\mathcal V_{\rm feas}^\varrho}

\newcommand{\ED}{\mathcal{SN}^\downarrow}

\newcommand{\EDV}{\boldsymbol{\mathcal{SN}}^\downarrow}

\newcommand{\elsmall}{\overset{{\rm el}}{\le}}

\usepackage{cleveref}
\crefname{equation}{Eq.}{Eqs.}
\crefname{observation}{Obs.}{Obs.}
\crefname{corollary}{Corollary}{Corollaries}
\crefname{lemma}{Lemma}{Lemmata}
\crefname{proof}{Proof}{Proofs}
\creflabelformat{proof}{#2proof#3}
\crefname{remark}{Remark}{Remarks}
\crefname{prop}{Proposition}{Propositions}

\begin{document}

\title{Characterizing high-dimensional multipartite entanglement beyond Greenberger-Horne-Zeilinger fidelities}

\author{Shuheng Liu}
\orcid{0000-0001-7130-1888}
\affiliation{State Key Laboratory for Mesoscopic Physics, School of Physics, Frontiers Science Center for Nano-optoelectronics, Peking University, Beijing 100871, China}
\affiliation{Vienna Center for Quantum Science and Technology, Atominstitut, TU Wien,  1020 Vienna, Austria}

\author{Qiongyi He}
\orcid{0000-0002-2408-4320}
\email{qiongyihe@pku.edu.cn}
\affiliation{State Key Laboratory for Mesoscopic Physics, School of Physics, Frontiers Science Center for Nano-optoelectronics, Peking University, Beijing 100871, China}
\affiliation{Collaborative Innovation Center of Extreme Optics, Shanxi University, Taiyuan, Shanxi 030006, China}
\affiliation{Hefei National Laboratory, Hefei 230088, China}

\author{Marcus Huber}
\orcid{0000-0003-1985-4623}
\email{marcus.huber@tuwien.ac.at}
\affiliation{Vienna Center for Quantum Science and Technology, Atominstitut, TU Wien,  1020 Vienna, Austria}
\affiliation{Institute for Quantum Optics and Quantum Information (IQOQI), Austrian Academy of Sciences, 1090 Vienna, Austria}

\author{Giuseppe Vitagliano}
\orcid{0000-0002-5563-3222}
\email{giuseppe.vitagliano@tuwien.ac.at}
\affiliation{Vienna Center for Quantum Science and Technology, Atominstitut, TU Wien,  1020 Vienna, Austria}

\begin{abstract}
Characterizing entanglement of systems composed of multiple particles is a very complex problem that is attracting increasing attention across different disciplines related to quantum physics. The task becomes even more complex when the particles have many accessible levels, i.e., they are of high dimension, which leads to a potentially high-dimensional multipartite entangled state. These are important resources for an ever-increasing number of tasks, especially when a network of parties needs to share highly entangled states, e.g., for communicating more efficiently and securely. For these applications, as well as for purely theoretical arguments, it is important to be able to certify both the high-dimensional and the genuine multipartite nature of entangled states, possibly based on simple measurements. Here we derive a novel method that achieves this and improves over typical entanglement witnesses like the fidelity with respect to states of a Greenberger-Horne-Zeilinger (GHZ) form, without needing more complex measurements. 
We test our condition on paradigmatic classes of high-dimensional multipartite entangled states like imperfect GHZ states with random noise, as well as on purely randomly chosen ones and find that, in comparison with other available criteria our method provides a significant advantage and is often also simpler to evaluate.
\end{abstract}

\maketitle

\section{Introduction}

Quantification of entanglement is a topic of wide interest crossing various areas of physics, because of its foundational interest
and also because it lies at the heart of quantum information theory with applications potentially in all areas of science~\cite{HorodeckiEntanglementReview2009,amico08,GuehneToth2009,Laflorencie16,FriisNatPhys19,RaussendorfComputer2001}. 
On the one hand, quantifying entanglement helps in navigating the complexity of many-body states~\cite{amico08,Laflorencie16} and is a resource for an ever increasing number of tasks, ranging from the simulation 
of many-body physics~\cite{amico08,Laflorencie16}, to quantum communication~\cite{ekert1991quantum}, sensing~\cite{toth2014quantum} and universal computation~\cite{jozsa2003role} among other things. On the other hand, it is also an extremely complex task, and reaches a daunting complexity especially when the system is composed of many particles with multiple accessible levels. 
Nevertheless, high-dimensional multipartite entangled states are attracting increasing attention also for applications, like quantum teleportation~\cite{LuoQuantum2019}, error-correction~\cite{ScottPRA2004,ArnaudCerf2013,GoyenechePRA2014}, quantum communication~\cite{malik2016multi}, violation of multipartite Bell inequalities~\cite{PhysRevLett.88.040404,RyuGreenberger2013,RyuMultisetting2014,LawrenceRotational2014,TangMultisetting2017}, randomness generation~\cite{JianweiMultidimensional2018} and Quantum Key-Distribution (QKD)~\cite{PhysRevLett.88.127902,SimonExperimental2006,EckerOvercoming2019,PhysRevLett.98.060503,VertesiClosing2010}. In fact, generating high-dimensional multipartite entangled states, especially the renowned Greenberger-Horne-Zeilinger (GHZ) states, has been the goal of numerous recent experiments, particularly with photonic systems~\cite{malik2016multi,LinMetalens2020,chi2022programmable,erhard2018experimental,LuoQuantum2019,PhysRevA.97.062309,MarioGeneration2014,dada2011experimental,ChristophIntegrated2015,JianweiMultidimensional2018,PhysRevLett.118.110501,PhysRevLett.120.030401,PanPRL18Qubit2018,PhysRevLett.123.170402,QuPhotonic2022,mair2001entanglement,PhysRevLett.128.240402,LiTwoMeasurement2023,HuOptimized2021,PhysRevLett.127.110505,ZhouQuantum2015,HuEfficient2020,valencia2020unscrambling,EckerOvercoming2019,bavaresco2018measurements,fickler2014interface,zhang2017simultaneous,ZhuIs2021,YunMultichip2023,HiesmayrObservation2016,bao2023very,Bouwmeester1999,pan2000experimental,Zhaoetal2003,lu2007experimental,gao2010experimental,CerveraExperimental2022}. 

Technically, the unambiguous quantification of entanglement is done via so-called {\it entanglement monotones} and is very challenging~\cite{Eltschka_2014}, as such measures are often defined from an average or a worst-case over all pure-state decompositions, which makes entanglement certification NP-hard \cite{GurvitsClassical2003}.
Furthermore, in a multipartite system all the difficulties of characterizing entanglement monotones are exacerbated, as each pure state decomposition element might now feature a different factorisation into parties. 
Because of that when it comes to the quantification of the genuine multipartite nature of the entanglement, it is important to consider bipartite entanglement monotones across the full set of bipartitions. 
In the literature, several distinct approaches have been developed to deal with the quantification of entanglement in both the high-dimensional and multipartite regimes. As we mentioned,
in the case of mixed states the space of quantum states becomes extremely rich and complex and it is complicated to even extend the definitions that are valid for pure states. In the latter case,   
one can consider bipartite entanglement monotones, like entanglement entropies, across the full set of bipartitions,
which can be enumerated with an index $\alpha$ going from $1$ to $\mathcal N = 2^{N-1}-1$ for a $N$-particle system. 
Starting from there, the extension of such a vector to mixed states and the definition of meaningful entanglement monotones can be done in multiple ways, that can capture some aspects of the high-dimensional as well as the multipartite nature of the entanglement~\cite{HuberStructure2013,HuberPerarnaudeVicentePRA13,KlocklCharacterizing2015,GuehneToth2009,FriisNatPhys19,Cobucci_2024}. However, in order to fully capture both aspects at the same time in a strict sense, one approach that has been developed is based on first ordering the entanglement entropy vector,
for example non-increasingly, and only then take the element-wise convex-roof extension to mixed states~\cite{HuberStructure2013,HuberPerarnaudeVicentePRA13} (see also \cite{Cobucci_2024}). 

\begin{figure}[h]
\centering
\includegraphics[width=0.9\textwidth]{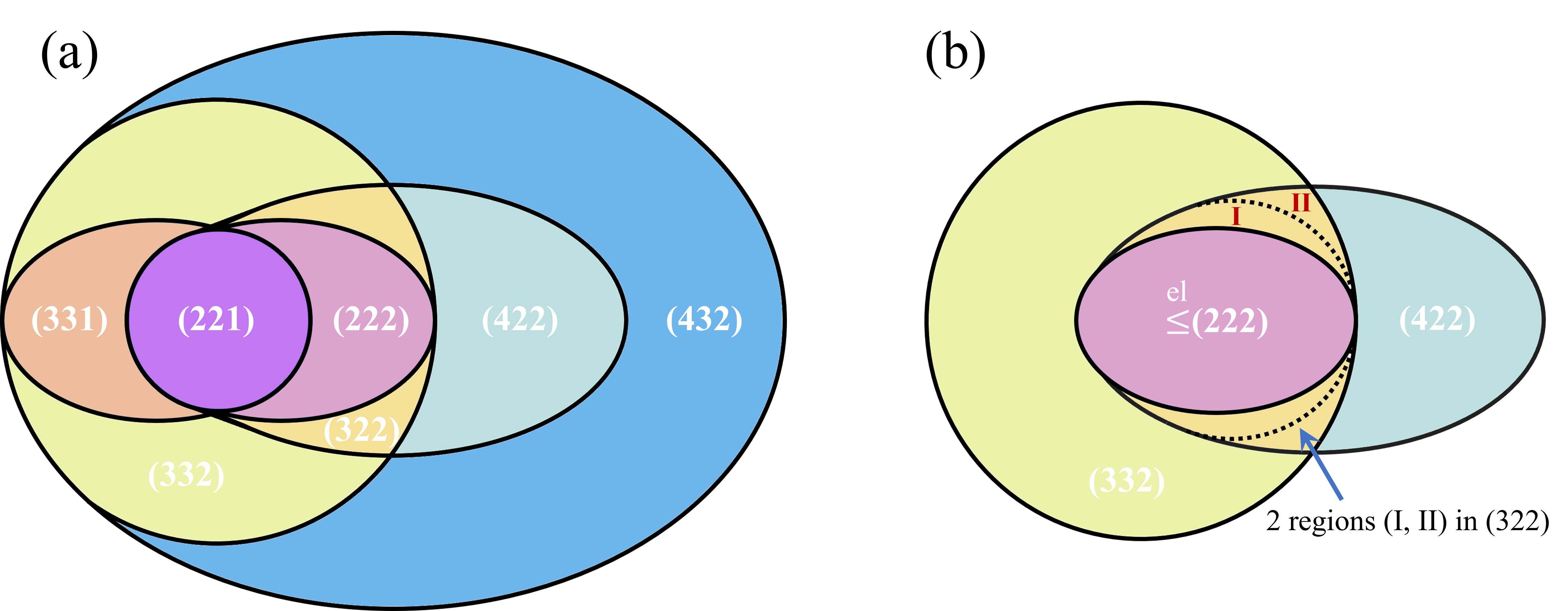}
\caption{(a) The structure of all possible Schmidt number vectors in a $4\times 3\times 2$ state space. Ellipses are only for illustration as the sets are not convex. 
Moreover, the extremal states have a complicated characterization and the structure is only partially nested. This is demonstrated more clearly in (b). (b) Taking vector $(322)$ as an example, it lies in the region covered by both $\vec{v}\elsmall (422)$ and $\vec{v}\elsmall (332)$, but outside $\vec{v}\elsmall (222)$. The regions (I) and (II) require careful distinction. Region (I) represents $(322)$ states that can be formed by mixtures of pure states with $\vec{v} \elsmall (322)$, while region (II) represents $(322)$ mixed states that cannot. 
For example, there can be states that only have decompositions with pure states with $\vec{v}= (422)$ or $\vec{v}= (332)$.
}
\label{fig:SNVectorStructure}
\end{figure}

In this work, we focus on the Rényi zero-entropy case, called {\it Schmidt number}~\cite{TerhalHorodeckiSchmidtNumber2000,SanperaBrussLewensteinPRA2001}, which, being discrete, is also a special case that allows for a better classification of the state as a resource as opposed to continuous measures~\cite{VidalEfficient2003,Van_den_Nest_2013,HuberStructure2013,HuberPerarnaudeVicentePRA13}. 
Working in a multipartite scenario we then consider the vector of Schmidt numbers across all possible bipartitions, called {\it entanglement dimensionality vector}~\cite{HuberStructure2013,HuberPerarnaudeVicentePRA13}.
Each element of such a vector is per s\'e already an entanglement monotone that is typically not easy to bound. 
Furthermore, Schmidt numbers across different bipartitions satisfy highly nontrivial relations amongst each other, even for pure states~\cite{JoshInequalities2014}.
This definition creates a structure of states with different entanglement-dimensionality vectors which is far more complex than that arising in the bipartite case, especially because it is not simply formed by nested convex sets. 
In particular, it is not possible to order all the different entanglement-dimensionality vectors, and states having a particular such vector do not form a convex set. 
Nevertheless, it is still possible to find a partially nested structure (cf. \cref{fig:SNVectorStructure}) and one can try to exclude that the state belongs to some of these sets. In particular, from the element-wise definition one gets nested
sets composed by vectors that satisfy an elementwise ordering $\vec v_1 \elsmall \vec v_2$, with which we denote the property that $\vec v_1$ has all the elements smaller or equal than $\vec v_2$. 
Because of this structure, usual methods to witness the Schmidt number become much more complex when extended to the multipartite case, and there is a lack of efficient methods to characterize the full entanglement-dimensionality vector in general~\cite{Huber_2010,HuberStructure2013,HuberPerarnaudeVicentePRA13}. 
It would be thus highly desirable to provide new approaches, especially some which avoid complex numerical optimizations.

This is encouraged by the fact that in recent years advancements in quantum state preparation have expanded the achievable Schmidt number vectors, which is often characterized through fidelities with GHZ-like states of high-dimensions~\cite{malik2016multi,erhard2018experimental,CerveraExperimental2022,hu2020experimental,bao2023very}, the latter being the target of the experiments. However, this approach has also some theoretical weaknesses: (i) first of all, it works mainly for such specific classes of well-aligned target states, (ii) furthermore, it only allows one to bound the smallest element of the entanglement-dimensionality vector. It remains thus challenging to extend this approach beyond current demonstrations, allowing for a more systematic investigation of generic quantum states. In this paper we address this issue, constructing a novel method to characterize the entanglement-dimensionality vector with measurements that have similar complexity as GHZ-like fidelities but allow to detect a wider and more generic set of states, also characterizing the entire vector rather than just one of its elements.

\section{Methods}\label{sec:methods}

\subsection{Definition of Schmidt-number vector}

To completely characterize entanglement across a $N$-partite state with high-dimensional constituents, it is useful to consider entanglement monotones, e.g., entropies 
across all $\mathcal N=2^{N-1}-1$ bipartitions.
Such entanglement entropies are defined first for pure states and for a given bipartition $(\alpha | \bar \alpha)$ by taking the entropy of the single-party marginal:
\be
S_\alpha(\psi):= S(\tr_{\bar \alpha}(\ketbra{\psi})) ,
\ee
where $S(\varrho)$ is a given entropy, which we take to be the zero entropy $S(\varrho) = \log({\rm rank}(\varrho))$. 

At this point, in principle one could consider any subset of these entropies across bipartitions and extend the definition to mixed states, for example via an average or worst-case over all decompositions. Or alternatively, one could consider, e.g., a norm of the vector having elements some (or all) these entropies. These would quantify, for example, how
much entanglement there is across a fixed bipartition or a specific subset of them. However, in the case of mixed states that are multipartite with high-dimensional constituents, 
capturing the genuine high-dimensional and multipartite nature of the entanglement is more subtle. This is because 
different decompositions might feature different entanglement dimensionalities across different bipartitions, and there is thus a priori no obvious canonical decomposition that one can rely on.
Hence, one idea that allows to account for the different possible factorisations of the different pure-state decompositions is to list all given entropies for all bipartitions and then order them non-increasingly, before
making an element-wise 
extension to mixed states~\cite{HuberStructure2013,HuberPerarnaudeVicentePRA13}.
This way, one defines measures like 
\be\label{eq:Eofalphagen}
\mathcal E^\downarrow_\alpha(\varrho) = \inf_{\mathcal D(\varrho)} \sum_k p_k S^\downarrow_\alpha(\psi_k) ,
\ee
where $S^\downarrow_\alpha$ are the components of the {\it ordered} vector of entropies of the marginal states for all bipartitions and the infimum is taken over all pure state decompositions $\mathcal D(\varrho) = \{p_k , \ket{\psi_k} \} : \varrho = \sum_k p_k \ketbra{\psi_k}$. Among all possible entanglement entropy vectors, only the von Neumann and the zero-entropy cases feature a nontrivial asymptotic structure that can be constrained by linear inequalities \cite{lieb1973fundamental,LiebRuskaiJMP73,JoshInequalities2014}.
In the latter case one defines for pure states the {\it Schmidt-rank vector} as the vector of Schmidt ranks with respect to all possible bipartitions
\begin{equation}
\SNV\left(\ket{\psi}\right)=\left(s_{1},s_{2},\cdots, s_{\mathcal N}\right) ,
\end{equation}
and then orders its elements non-increasingly. As we mentioned, the ordering becomes crucial for extending this notion to the multipartite case and for clarity we denote the ordered vectors by $\EDV$.

Then, for mixed states we define the entanglement-dimensionality or Schmidt-number (SN) vector element-wise via a worst case over all pure state decompositions, i.e., 
\be\label{eq:SNVdef}
\ED_\alpha(\varrho)= \inf_{\mathcal D(\varrho)} \max_{\ket{\psi_k} \in \mathcal D(\varrho)}  \ED_\alpha(\ket{\psi_k}) ,
\ee
with again the crucial requirement that the pure-state Schmidt-rank vectors are ordered non-increasingly.

We note how this definition makes it a priori quite challenging to determine the entire entanglement-dimensionality vector, as in principle one should consider all possible pure-state decompositions in a very complex way, there being in principle a {\it different optimal} decomposition for each element of the SN vector. 
This is because, as we mentioned, each pure-state decomposition can feature in principle pure states with different Schmidt ranks across different bipartitions, and one has to order them non-increasingly.
Let us illustrate this difficulty via an example in a three-partite state in a $(4\times 3 \times 2)$-dimensional system~\cite{HuberStructure2013}. In principle there could exist a density matrix that features decompositions with pure states having ordered
Schmidt number vectors that are either $\vec v = (422)$ or $\vec v = (332)$. That is, we can decompose such a density matrix as either $\varrho = \sum_k p_k \ketbra{\psi_k^{(422)}}$ or $\varrho = \sum_k p_k' \ketbra{\psi_k^{(332)}}$, where we denoted with a superscript the {\it ordered} Schmidt-rank vector of the pure states. In such a situation, the Schmidt-number vector according to the definition given in \cref{eq:SNVdef} is given by $\vec v(\varrho) = (322)$, as one has to take the {\it element-wise infimum} across all possible decompositions, and we have imagined a situation in which both decompositions of the two above cases exist. Nevertheless, the existence of these two decomposition alone does not guarantee the existence of a decomposition of $\varrho$ with pure states having $\EDV = (322)$, which in principle may thus not exist~\footnote{Clearly, this and analogous example look like quite special and not necessarily existing cases. In fact, so far no explicit example of such a situation exists. However, we cannot exclude the existence of such cases either.}. 
See also \cref{fig:SNVectorStructure} for an illustration. This makes it also harder to extend pure-state witnesses of the Schmidt rank to exclude a given entanglement-dimensionality vector for mixed states, as we are going to observe in the following.  

We stress the importance of this definition: a direct definition based on pure-state entanglement-dimensionality vectors would imply that the above mentioned class of states do not have a well-defined entanglement-dimensionality vector, since it simultaneously admits (422) and (332) decompositions. Because of that, avoiding the prior ordering of the SN vector, could lead to a definition which is not well-defined for arbitrary states.

\subsection{Known witnesses of entanglement-dimensionality vector}\label{sec:knownSNwit}
The most typical witnesses of the entanglement-dimensionality, that can be extended to characterize the SN vector are given in terms of fidelities with respect to pure target states~\cite{malik2016multi,Cobucci_2024}.
Concretely, given a non-increasingly ordered entanglement-dimensionality vector $\vec v := (v_1,\dots , v_{\mathcal N})$ and a target pure state $\ket{\Psi}$ one looks for its maximal fidelity with pure states having given (ordered) entanglement dimensionality vectors. Then one collects the following values
\be\label{eq:fidmaxmulti}
F_{\rm max}(\vec v , \Psi) := 
\max_{\EDV(\phi) \elsmall \vec v} |\langle \phi | \Psi \rangle|^2  ,
\ee
which can be used to divide the possible (candidate SN) vectors $\vec v$ into {\it feasible} and {\it unfeasible} as follows.

Given an arbitrary mixed state $\varrho$ and its fidelity $F(\varrho, \Psi)$ with a target state $|\Psi\rangle$, the feasible vectors are defined as:
\be
\Vfeas:=\{\vec{v} \mid F(\varrho,\Psi) \leq 
F_{\rm max}(\vec v , \Psi)\} .
\ee
Contrarily, the unfeasible vectors are defined as those such that $F(\varrho,\Psi) > F_{\rm max}(\vec v , \Psi)$. Once the set of all possible vectors is divided into these two mutually exclusive subsets, we can establish that each element of the entanglement-dimensionality vector $\EDV(\varrho)$ must be no less than the minimum value of the corresponding component in the set $\Vfeas$, namely
\be\label{eq:feasiblesetDef}
\ED_\alpha(\varrho) \geq \min_{\vec{v} \in \Vfeas} v_\alpha, \quad \forall \alpha.
\ee
This general strategy based on characterizing feasible versus unfeasible vectors can be used in many cases to extend pure-state bounds for given Schmidt-rank vectors to the more general case of mixed states, still taking into account all complications arising from the different possible structures in the different decompositions, that we discussed earlier.

Note once more however, that one cannot generally conclude that $\varrho$ has a particular entanglement-dimensionality vector which is ``greater'' than $\vec v$, because such an ordering is lost in the multipartite case. Furthermore, even though the fidelity is a convex function of the state, the characterization of the Schmidt-number vector via fidelities becomes much more complex than in the bipartite case. Moreover, even if the characterization can still be made via just pure-state fidelities, one still needs to know the Schmidt eigenvalues of the target state across all bipartitions and, even after that, perform in general a quite demanding optimization. See \cref{Appendix:KnownWitnesses} for more details. 
This is because, as we observed, one needs to consider pure states with many different Schmidt-number vectors.

Thus, often it is necessary to restrict the attention to a simplified version of the problem, namely exclude only entanglement dimensionality vectors that feature a given minimal element~\cite{malik2016multi,FriisNatPhys19,Cobucci_2024}.
For this simplified task it is common to choose 
a target state which is highly symmetric, a typical example being the (high-dimensional) Greenberger-Horne-Zeilinger (GHZ) state:
\be\label{eq:ghzpurestate}
\ket{\Psi^d_{\rm GHZ}} = \frac{1}{\sqrt{d}} \sum_{i=0}^{d-1} \ket{i}^{\otimes N} .
\ee
This then leads to additional simplification due to the fact that the state is invariant under permutation of the parties and the one-body marginals are all maximally mixed. 

Because of that, with such a target state the bound $F_{\rm max}(\vec v , \Psi^d_{\rm GHZ})$ in fact coincides for all entanglement-dimensionality vectors $\vec v$ that have a given minimal element $v_{\mathcal N}$. 
In general this is one of the few examples for which the bound in \cref{eq:fidmaxmulti} 
is known and can thus be applied to practical entanglement detection.
Beyond fidelities, a few further, also nonlinear witnesses that can be used to characterize the entanglement-dimensionality vector to some extent have been derived. 
In particular, in Refs.~\cite{HuberStructure2013,KlocklCharacterizing2015}.
The first method relies on measurements on a given multipartite vector basis $\ket{\eta}$, where $\eta=(i_1,\dots, i_N)$ is a multi-index with $N$ entries, taking values from $0$ to $d_n-1$, where $d_n$ is the dimension of particle $n$. Concretely, one finds a lower bound to the linear entropy vector elements
\begin{equation}\label{eq:boundlinentvec}
(\mathcal E^\downarrow_k)^{\rm lin}(\varrho) \geq \mathcal B\left(\bra{\eta}\varrho \ket{\eta^\prime} \right) , 
\end{equation}
in terms of matrix elements $\varrho_{\eta \eta^\prime}:=\bra{\eta}\varrho \ket{\eta^\prime}$ for suitably chosen subsets of pairs of indices $(\eta, \eta^\prime)$.
Here the expression of the bound (which is non-linear in the matrix elements $\varrho_{\eta \eta^\prime}$) is omitted for simplicity and is given in the~\cref{Appendix:KnownWitnesses}.
In turn \cref{eq:boundlinentvec} can be translated into a bound on the entanglement-dimensionality vector elements by using the relation
\be\label{eq:boundSNfromLEV}
\ED_k(\varrho) \geq \left\lceil 2 / \left(2- ((\mathcal E^\downarrow_k)^{\rm lin}(\varrho))^2 \right) \right\rceil ,
\ee
where $\lceil x \rceil$ is the usual ceil function. 

The second method is instead invariant under any particular basis expansion of the density matrix, but works only for witnessing the ranks of the single-particle marginals~\cite{KlocklCharacterizing2015}, which do not exactly coincide with any of the elements of the SN vector as per the definition in \cref{eq:SNVdef}. 
Here we call it correlation-tensor-norm criterion and works as follows (a more detailed description is given in~\cref{Appendix:KnownWitnesses}).

Let us consider an $N$-partite density matrix with local dimensions $d_n$ and expand it in terms of single-particle orthonormal bases:
\begin{equation}
\varrho=
\sum_{\mu_1, \ldots, \mu_N} \av{g^{(1)}_{\mu_1} \otimes \cdots \otimes g^{(N)}_{\mu_N}} g^{(1)}_{\mu_1} \otimes \cdots \otimes g^{(N)}_{\mu_N},
\end{equation}
where $0\leq \mu_n\leq d_n^2-1$. Specifically, consider single-particle bases $g^{(n)}_{\mu_n}$ that are composed of the identity matrix $g^{(n)}_0=\id_{d_n}/\sqrt{d_n}$ and the (normalized) generators of the $su(d_n)$ algebra $\{\sigma^{(n)}_{1} , \dots , \sigma^{(n)}_{d_n^2-1} \}$.
In particular when all $d_n$ are equal to $d$, in \cite{KlocklCharacterizing2015} the authors considered the correlations of $su(d)$ observables among two or more of the particles, and defined the quantity
\be
\mathcal{C}_2\left(\varrho\right) := \sum_{m=2}^N \sum_{|\alpha|=m} \left\| T^{(\alpha)} \right\|_2^2 ,
\ee
where 
$T^{(\alpha)}$ is the $|\alpha|$-body correlation tensor:
\be
T_{\mu_1, \mu_2, \ldots, \mu_{|\alpha|}}^{(\alpha)} := \av{\sigma^{(n_1)}_{\mu_{n_1}} \otimes \cdots \otimes \sigma^{(n_{|\alpha|})}_{\mu_{n_{|\alpha|}}}} ,
\ee
where as usual $\alpha$ labels the partition $(\alpha | \bar \alpha)$, which is composed of $|\alpha|$ versus $N-|\alpha|$ particles and 
its $2$-norm is defined as
\be
\|T^{(\alpha)}\|_2:=\sqrt{\sum_{\mu_1, \mu_2, \ldots, \mu_{|\alpha|}} \left(T_{\mu_1, \mu_2, \ldots, \mu_{|\alpha|}}^{(\alpha)}\right)^2 } ,
\ee
where again the indices run from $1$ to $d^2-1$ for each particle.
For example, the $2$-particle correlation tensor norm of a tripartite state is 
\be
\mathcal{C}_2\left(\varrho\right)=\left\|T^{(12)}\right\|_2^2+\left\|T^{(23)}\right\|_2^2+\left\|T^{(13)}\right\|_2^2+\left\|T^{(123)}\right\|_2^2 . 
\ee

The criterion in \cite{KlocklCharacterizing2015} gives a necessary condition that must hold for every density matrix $\varrho$ that is a convex combination of pure states with Schmidt ranks across bipartitions with $|\alpha|=1$ given by $\left(k_1, k_2, \ldots, k_N\right)$. This necessary condition is given by the inequality
\begin{equation}\label{eq:corrtensineq}
\mathcal{C}_2\left(\varrho \right) \leqslant d^N+N-1-\sum_n \frac{d}{k_n} .
\end{equation} 
As a consequence, violating \cref{eq:corrtensineq} indicates that at least one of the Schmidt numbers of $\varrho$ across bipartitions with one particle versus the rest is greater than the corresponding rank in the vector $\left(k_1, k_2, \ldots, k_N\right)$. This vector is in general different from the Schmidt number vector as we have considered it, even though it still provides
information about the dimensionality of entanglement in the state. 

These are essentially all criteria that exist to characterize the entanglement dimensionality of a multipartite quantum state.
It is worth mentioning that many more approaches exist for detecting various forms of entanglement in the bipartite or multipartite case~\cite{HorodeckiEntanglementReview2009,GuehneToth2009,FriisNatPhys19}, including witnesses and nonlinear criteria~\cite{GuehneLutkenhaus06,ZhuTeoEnglert2010}. It would be an interesting, although still potentially challenging question, to extend them to characterize (partly or fully) the entanglement-dimensionality vector of a multipartite system. Below we present an approach that does so for the covariance matrix criterion~\cite{Liu2024bounding,Guhne2004Characterizing,guhnecova,gittsovich08,GittsovichPRA10}.

\section{Results}

\subsection{New method to characterize the Schmidt-number vector}

We start from a recently developed method to witness the Schmidt number across bipartitions from the covariance matrix of local orthonormal operator bases~\cite{Liu2024bounding,Guhne2004Characterizing,guhnecova,gittsovich08,GittsovichPRA10}~\footnote{Note that the same approach would work even when the operator basis is not Hermitian. However we illustrate here our method using a Hermitian basis, as it is most typically considered in the literature, and also for simplicity of notation.}.
Here, working explicitly in a multipartite setting, we consider single-particle operators $\{ g_\mu^{(n)} \}_{\mu=1}^{d_n^2}$ where $n$ labels the particles and $d_n$ are their dimensions. We consider basis operators ortho-normalized as $\tr( (g_\mu^{(n)})^\dagger g_\nu^{(n)})=\delta_{\mu \nu}$. Given a bipartition $(\alpha | \bar \alpha)$ we can construct basis for party $\alpha$ by taking tensor products of single-particle operators among all the particles in $\alpha$: $\{g^{(\alpha)}_{K}\} = \{ g_\mu^{(n)} \otimes  g_\nu^{(m)} \otimes \dots \}_{n,m\dots \in \alpha}$. To be more clear and compact, we used a single capitalized multi-index $K=(\mu \nu \dots..)$ for such a basis.

First, let us fix a bipartition $\alpha$ and consider the cross-covariances
\be
[\C^{(\alpha)}]_{K L} = \av{g^{(\alpha)}_{K} \otimes g^{(\bar\alpha)}_{L}}_\varrho - \av{g^{(\alpha)}_{K}}_\varrho \av{g^{(\bar\alpha)}_{L}}_\varrho
\ee
on a quantum state $\varrho$.
Using the results of Ref.~\cite{Liu2024bounding} (generalizing those of Refs.~\cite{guhnecova,gittsovich08,GittsovichPRA10}) we know that all states $\varrho$ with a Schmidt number at most equal to $r_\alpha$ across bipartition $\alpha$ have to satisfy
\be
f_\alpha(\varrho):= \tr|\C^{(\alpha)}| - \sqrt{[1 - \tr(\varrho^2_\alpha)][1 -\tr(\varrho^2_{\bar \alpha})] } +1 \leq r_\alpha , \label{eq:cor1} 
\ee
where $\tr|X|=\tr\sqrt{X^\dagger X}$ is the trace norm of a matrix and we named the left-hand side as $f_\alpha(\varrho)$ to shorten the notation in the following. 
Note that the left-hand side of \cref{eq:cor1} is invariant under changes of bases for the parties $(\alpha | \bar \alpha)$.
We sketch the derivation of this statement in the~\cref{Appendix:CMCRecall} and refer to \cite{Liu2024bounding} for the more complete proof.

Now, let us try to extend such a relation to identify the entire entanglement-dimensionality vector.
For that, we need to consider all bipartitions at the same time, and correspondingly all matrices $\C^{(\alpha)}$.
Then, we also need to consider all potential vectors $\vec v$ and first identify the set of feasible ones, as in the case of the fidelity-based witnesses.
In our case, such a feasible set is given by the vectors
$\vec v=(v_1 , v_2 , \dots , v_{\mathcal N})$ that are such that the following system of inequalities is satisfied:
    \be\label{eq:obs2sys}
     \left\{ \begin{array}{cc}
            f_1(\varrho)  &\leq R_1 , \\   
            f_2(\varrho) &\leq R_2 , \\   
             \vdots & \\ 
            f_{\mathcal N}(\varrho) &\leq R_{\mathcal N} ,  
    \end{array}
    \right.
    \ee
    where $\vec R = (R_1 , \dots , R_{\mathcal N})$ is a vector of real numbers such that 
    \be
    \vec R \prec \vec v ,
    \ee
    which means that 
    \be
     \begin{aligned}
         \sum_{k=1}^K R^\downarrow_k &\leq \sum_{k=1}^K v_k \quad \text{for all} \quad K \leq \mathcal N , \\
         \sum_{k=1}^{\mathcal N} R^\downarrow_k &= \sum_{k=1}^{\mathcal N} v_k ,
     \end{aligned}
    \ee
    where $R^\downarrow$ are the elements of $\vec R$ ordered non-increasingly. 
To be more explicit, we note that in this case, a vector $\vec v \in \Vfeas$ is feasible relative to a quantum state $\varrho$ whenever there exists a decomposition of the form
\be\label{eq:decompoSNV}
\varrho = \sum_k p_k \varrho_{\vec s_k} ,
\ee
where $p_k$ are probabilities and $\varrho_{\vec s_k}$ are pure states with a given (unordered) Schmidt number vector $\vec s_k$ that is such that $(s^\downarrow_k)_j \leq v_j$ when the components of $\vec s_k$ are ordered non-increasingly. 
This system of inequalities \cref{eq:obs2sys} can be solved with a linear program in the variables $R_\alpha$. If some $\vec R$ solution to this problem can be found, then we have $\vec v \in \Vfeas$, otherwise $\vec v \notin \Vfeas$. 
As described earlier for the case of fidelity witnesses, we can then characterize the allowed $\EDV(\varrho)$ from the feasible set from \cref{eq:feasiblesetDef}.
Note that criteria obtained with this approach (including also the fidelity method) become more stringent, the more vectors are excluded from the set of feasible ones.
We have already mentioned that in practice the fidelity witnesses that are employed (i.e., those with respect to the GHZ states) are only able to distinguish vectors with different smallest elements.
Here, as compared to fidelity witnesses, we provide an explicit method that can be practically applied to exclude many more vectors from the feasible set, e.g., not just those that have a given minimal element.

\subsection{Recovering GHZ-like fidelity witnesses}

Simpler conditions can be obtained from this system of inequalities, that already provide important complementary information with respect to current methods.
For example, by taking the maximum of the left-hand side of \cref{eq:obs2sys} we obtain 
\be
\max_\alpha f_\alpha(\varrho) 
\leq \ED_1 (\varrho) \leq v_1
\ee
for every $\boldsymbol{v} \in \mathcal{V}_{\text{feas}}^{\varrho}$, which allows one to bound the {\it maximal} entanglement dimensionality across all bipartitions. This information, in turn, complement a bound on the minimal element of $\EDV(\varrho)$ and, for example, would allow a better estimate of the minimal dimensionality-cost of reproducing the state's correlations.
Furthermore, the scaling of the largest component $\ED_1 (\psi)$ of the entanglement-dimensionality vector effectively lower-bounds the possibility of exponential quantum speed-ups in the pure-state $N$-qubit circuit model. If $\ED_1 (\psi)$ grows only such that $\log \ED_1 (\psi)=O(\log N)$ throughout the computation, then the computation admits an efficient classical simulation \cite{VidalEfficient2003,VandenNestUniversal2013}.

Let us now observe how to bound the minimal element of $\EDV(\varrho)$ also from our approach, recovering the GHZ-type fidelity witness. First, we note that it is possible
to further bound the left-hand side of \cref{eq:cor1}.
Consider orthonormal bases of the single particle operator space for each particle $\{ g^{(n)}_\mu \}_{\mu=1}^{d_n^2}$ and let $d = \min_n d_n$ be the smallest among the single-particle dimensions. From the fact that 
\be\label{eq:boundfromftolin}
\sum_{\mu=1}^{d^2} \av{g^{(1)}_{\mu}\otimes \dots \otimes g^{(N)}_{\mu}}_\varrho \leq f_\alpha(\varrho)
\ee
holds for every $\alpha$ we get the following inequality
\be\label{eq:weakerthancor1} 
\sum_{\mu=1}^{d^2} \av{g^{(1)}_{\mu}\otimes \dots \otimes g^{(N)}_{\mu}}_\varrho \leq v_{\mathcal N} ,
\ee
that now must be satisfied by all states that are such that the last element of their entanglement-dimensionality vector is given by $v_{\mathcal N}$. Thus, a violation of this inequality implies that $\varrho$ has the minimal entanglement dimensionality which is larger than $v_{\mathcal N}$.
This condition simply follows from 
taking the minimum among all $\alpha$ in \cref{eq:cor1} and using the bound in \cref{eq:boundfromftolin}. 
Thus, as in the GHZ-fidelity-witness case, whenever \cref{eq:weakerthancor1} is satisfied, all vectors $\vec v$ with the smallest element equal to $v_{\mathcal N}$ are feasible. On the contrary, whenever \cref{eq:weakerthancor1} is violated for a certain $v_{\mathcal N}^\prime$, then all vectors $\vec v^\prime$ with the smallest element equal to it are excluded from the feasible set.

As a result, we get two independent conditions, \cref{eq:obs2sys,eq:weakerthancor1} that provide complementary information about the entanglement-dimensionality vector, as we will clarify afterwards with some examples. 
However, the second condition \eqref{eq:weakerthancor1}, and in particular its effectiveness for detecting the dimensionality of entanglement, depends on the concrete choice of the various single-particle bases. As we will also emphasize later, a good choice for such bases is not obvious for an arbitrary density matrix $\varrho$, and further optimization is potentially required for implementing it. In fact, such a condition has some relation with fidelity witnesses discussed earlier. To see this, let us consider the situation in which all particles have the same dimension and we make a canonical choice of the $\{ g^{(n)}_{\mu}\}$, which is the same for all $n$. 
Specifically, we choose $\{ g^{(n)}_{\mu}\}$ such that the expression
\be\label{psiNdstate}
\frac 1 d \sum_{\mu=1}^{d^2} g_\mu^{(1)} \otimes \dots \otimes g_\mu^{(N)} := \ketbra{\Psi}
\ee
defines a quantum state whose single-body marginals are all maximally mixed. 
The same will remain true after applying arbitrary local unitaries to the state, which corresponds to changing the local bases via such a unitary. The class of states with this property are called
$1$-uniform in the literature about multipartite entanglement classification~\cite{ScottPRA2004,ArnaudCerf2013,GoyenechePRA2014} and includes the canonical GHZ state.
A detailed proof of \cref{eq:weakerthancor1} and its relation with fidelities to $1$-uniform states can be found in the~\cref{Appendix:ProofWeakerthancor1}.

\subsection{Comparison with the other methods}

Thus, once the quantities in $f_\alpha(\varrho)$ are known for a given quantum state $\varrho$ for all bipartitions, one can combine both conditions in \cref{eq:obs2sys,eq:weakerthancor1} and obtain a criterion that is stronger than fidelity witnesses with respect to GHZ-type states. Note that the quantities needed for both conditions
are 
$k$-particle correlation terms of the form $\av{g_\mu^{(1)} \otimes \dots \otimes g_\mu^{(k)}}$. 
This is similar to what is needed for typical witnesses of the Schmidt number vector (e.g., fidelities to GHZ-like target states), that contain $N$-particle correlators like in \cref{eq:weakerthancor1}.

\setlength{\tabcolsep}{5pt}
\begin{table}[hb]
\caption{Entanglement-dimensionality vector detected by the various known methods on $10000$ randomly sampled quantum states in a $3\times 3 \times 3$ system. We indicate the percentage of the total states that are detected having every given entanglement-dimensionality vector. The vector $(331)$ is never detected and is thus omitted.}\label{table:1}
\begin{equation}\nonumber
\begin{array}{|c|c|c|c|c|c|c|}
\hline \mathcal{S N} & (111) & (221) & (222) & (322) & (332) & (333) \\
\hline \begin{array}{c} \text {correlation tensor}\\ \text {norm, \cref{eq:corrtensineq} 
}\end{array} & 89.6 \% & 8.2 \% & 0.9 \% & 0.9  \% & 0.4 \% & 0.0 \% \\
\hline \begin{array}{c} \text {linear entropy vector}\\ \text { \cref{eq:boundSNfromLEV} + \cref{eq:boundlinentvec} }\end{array} & 18.1 \% & 20.4 \% & 61.5 \% & 0.0 \% & 0.0 \% & 0.0 \% \\
\hline \text {\cref{eq:weakerthancor1}  } & 48.1 \% & 0.0 \% & 48.3 \% & 0.0  \% & 0.0 \% & 3.6 \% \\
\hline \text { \cref{eq:obs2sys} } & 29.0 \% & 55.1 \% & 0.0  \% & 0.0 \% & 11.6 \% & 4.3 \% \\
\hline \text { \cref{eq:obs2sys}+ \cref{eq:weakerthancor1} } & 29.0 \% & 19.1 \% & 21.5 \% & 14.5  \% & 10.5 \% & 5.4 \% \\
\hline
\end{array}
\end{equation}
\end{table}

Let us now illustrate more clearly some of the practical aspects of our method, and also compare it to other known witnesses of the entanglement-dimensionality vector.
First of all, it is clear that the combination of the two inequalities (\ref{eq:obs2sys},\ref{eq:weakerthancor1}) provides a witness which is stronger than the fidelity witness with respect to the corresponding $1$-uniform state \eqref{psiNdstate}. As we mentioned, while \cref{eq:obs2sys} does not require further optimization, one can try to look for the optimal bases to evaluate \cref{eq:weakerthancor1}, which would correspond to calculate a fidelity witness to an optimal $1$-uniform state that depends on $\varrho$, which however is a nontrivial task in general. See for example \cite{Cobucci_2024} or \cite{Zhang_2024} for recent approaches related to this problem.
With these premises in mind, let us apply the combination of criteria in \cref{eq:obs2sys,eq:weakerthancor1} to witness the Schmidt number vector of randomly sampled states. In practice, to make numerical calculations we considered a tripartite system of dimension $d_1=d_2=d_3=3$ and sampled 10000 states in a way such that highly entangled states are generated with a significant probability (see~\cref{Appendix:SampleStates}). We then tested the various methods available to detect the Schmidt number vector, which are discussed in \cref{sec:methods} and in \cref{Appendix:KnownWitnesses}, and obtained the result summarized in \cref{table:1}.

We can clearly see that our method outperforms the others, even already by using just \cref{eq:obs2sys}. We also see that considering \cref{eq:weakerthancor1} on top of it actually improves the detection significantly. For applying \cref{eq:weakerthancor1} as well as the criterion based on the linear entropy vector in \cref{eq:boundSNfromLEV,eq:boundlinentvec}
we considered a heuristic optimization over local bases, using standard built-in functions of numerical software. Thus, as we mentioned, in a given practical scenario there might be more room for further optimizing them.

A situation in which the optimal way of evaluating \cref{eq:weakerthancor1} and \cref{eq:boundSNfromLEV,eq:boundlinentvec} is clearer is given by states that are close to a given pure state. 
Therefore, we also tested our method in such a situation. First, we considered the canonical high-dimensional GHZ state mixed with random noise, and we found analogous results as in \cref{table:1}, i.e., that our method outperforms the others, and in particular improves over the fidelity witness alone. The results are summarized in \cref{table:2}. See also \cref{Appendix:SampleStates} for additional technical details regarding these numerical results.

\setlength{\tabcolsep}{5pt}
\begin{table}[hb]
\caption{Entanglement-dimensionality vector detected by the various known methods on $10000$ samples of quantum states of the form \eqref{eq:GHZwithrandomnoise} in a $3\times 3 \times 3$ system. We indicate the percentage of the total states that are detected having every given entanglement-dimensionality vector. The vector $(331)$ is never detected and is thus omitted.}\label{table:2}
\begin{equation}\nonumber
\begin{array}{|c|c|c|c|c|c|c|c|}
\hline \mathcal{S N} & (111) & (221) & (222) & (322) & (332) & (333) \\
\hline \begin{array}{c} \text {correlation tensor}\\ \text {norm, \cref{eq:corrtensineq} }\end{array} & 86.9 \% & 8.9 \% & 1.2 \% & 1.1 \% & 0.9 \% & 1.0 \% \\
\hline \begin{array}{c} \text {linear entropy vector}\\ \text { \cref{eq:boundSNfromLEV} + \cref{eq:boundlinentvec} }\end{array} & 9.8 \% & 15.2 \% & 62.4 \% & 1.1  \% & 0.9 \% & 10.6 \% \\
\hline \text { \cref{eq:weakerthancor1} } & 31.2 \% & 0.0 \% & 33.2 \% & 0.0 \% & 0.0 \% & 35.6 \% \\
\hline \text { \cref{eq:obs2sys} } & 24.9 \% & 50.6 \% & 0.0 \% & 0.0 \% & 12.2 \% & 12.3 \% \\
\hline \text { \cref{eq:obs2sys}+ \cref{eq:weakerthancor1} } & 24.9 \% & 6.3 \% & 32.4 \% & 0.8 \% & 0.0 \% & 35.6 \% \\
\hline
\end{array}
\end{equation}
\end{table}

\begin{figure}[h]
\centering
\includegraphics[width=0.75\textwidth]{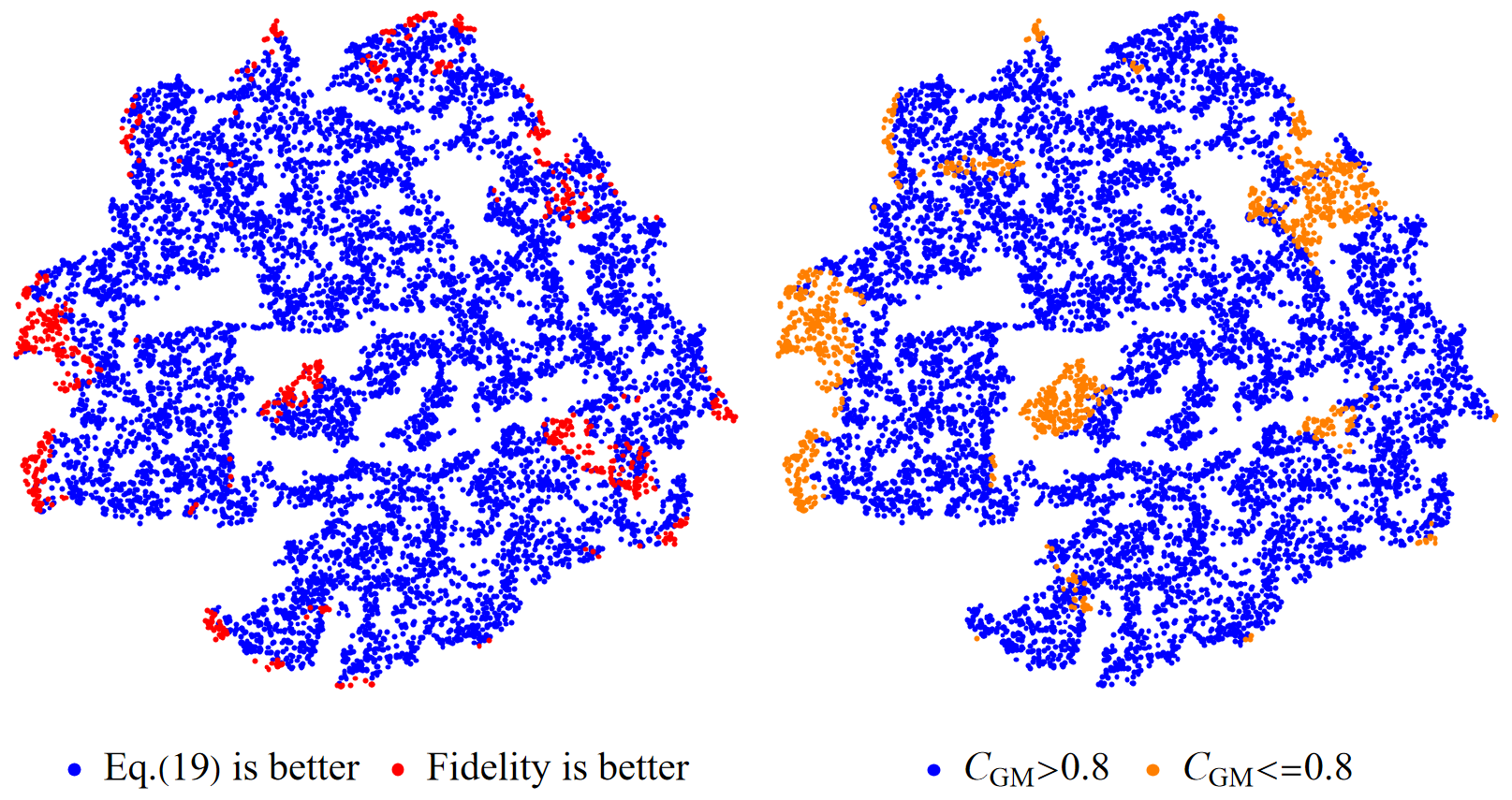}
\caption{
Comparison of fidelity-witness based criterion and \cref{eq:obs2sys} over random states of the form \cref{eq:varrho432}, visualized with a 2-dimensional clustering made using t-SNE.  (a) Blue points represent states detected with higher white-noise $(1-p)$ by our criterion \eqref{eq:obs2sys}, while red points correspond to states that are detected with a higher tolerance by the fidelity witness with respect to $\ket{\psi_{432}}(\vec c)$. (b) Blue points represent states for which $C_{\mathrm{GM}}>0.8$, while orange points correspond to those with $C_{\mathrm{GM}}\leq 0.8$.  }\label{fig:GMConcurrence}
\end{figure}

Second, we consider states close to randomly generated pure states with GHZ-type Schmidt number vector in a $(4\times 3\times 2)$-dimensional system. 
More precisely we consider states of the form
\begin{equation}\label{eq:432purestate}
\ket{\psi_{432}}(\vec c):=c_1 \ket{000}+c_2 \ket{111}+c_3 \ket{012} +c_4 \ket{123}
\end{equation}
where $\vec c = (c_1,c_2,c_3,c_4)$ is a unit vector of complex coefficients. The case with all $c_i = \tfrac{1}{2}$ was considered in ~\cite{HuberStructure2013} as a simple example to show the partial ordering structure of the entanglement-dimensionality vector. In our simulations we take uniformly sampled random coefficients $c_i$ and mix the state with white noise:
\begin{equation}\label{eq:varrho432}
\varrho(p,\vec c)=p\ketbra{\psi_{432}}(\vec c)+(1-p)\frac{\mathbb{1}}{24}.
\end{equation}
We try to detect $\varrho(p,\vec c)$ using \cref{eq:obs2sys} and the fidelity with respect to $\ket{\psi_{432}}$ for comparison. All other criteria mentioned earlier are worse than either our method or the fidelity with respect to $\ket{\psi_{432}}$.
To certify $\SNV(\ket{\psi_{432}})=(4,3,2)$, one must exclude $(4,2,2)$ and $(3,3,2)$ simultaneously, the two cases being not comparable (cf \cref{fig:SNVectorStructure}). 
We make this comparison between the two criteria across a total of 10000 samples and use t-distributed Stochastic Neighbor Embedding (t-SNE)~\cite{van2008visualizing} as a dimension reduction technique to show the data. This is a method that ensures that states that are close to each other, remain close also in the lower-dimensional projected figure. The results are shown in \cref{fig:GMConcurrence}. 

We find that our method outperforms the fidelity witness in most of the cases (about $92\%$). It is also worth noting that the points that are detected by our method mostly coincide with states that have a relatively high value of the so-called
genuine multipartite concurrence, which is defined as the lowest element of the entropy vector in \cref{eq:Eofalphagen} where the entropy of choice is the linear entropy~\cite{QiangGenuine2018,HashemiGenuinely2012,MaMeasure2011}:
\begin{equation}\label{eq:linentrvec}
C_{\rm GM}(\varrho):= (\mathcal E^\downarrow_{\mathcal N})^{\rm lin}(\varrho) =\inf_{\mathcal D(\varrho)} \sum_k p_k (\min_{\alpha} S^{\rm lin}_{\alpha}(\psi_k)) .
\end{equation}
However, notice that $C_{\rm GM}$ is not high enough to witness a $\SNV$ better than either our criterion \eqref{eq:obs2sys} or the fidelity with respect to $\ket{\psi_{432}}$.

\section{Conclusions and Outlook}

In conclusion, we have presented a new approach to find witnesses for the entanglement-dimensionality vector in multipartite systems, which leads to a criterion that is strictly stronger than fidelity witnesses with respect to $1$-uniform states such as the GHZ states. Such fidelity-based witnesses represent the most widely used ones in practice. We have shown that our method, while requiring similar measurements as commonly used witnesses, significantly improves over known methods on a wide class of states, which include important paradigmatic examples useful for applications. Further developments of our approach can be obtained by extending known bipartite entanglement witnesses to detect the Schmidt number or finding new ones, and then extending them to a multipartite scenario. In particular, one
can try to further exploit general approaches that would encompass a wide range of witnesses to optimize over~\cite{HorodeckiEntanglementReview2009,GuehneToth2009,FriisNatPhys19}, also nonlinear ones such as the covariance matrix criterion~\cite{Liu2024bounding} and others~\cite{GuehneLutkenhaus06,ZhuTeoEnglert2010,HuberStructure2013}. This would hence represent a promising direction for further research in this topic.

\section{Acknowledgments} 
We thank Jordi Tura for discussions. This work is supported by the National Natural Science Foundation of China (Grants No. 12125402, 12534016, No. 12405005), Beijing Natural Science Foundation (Grant No. Z240007), and Quantum Science and Technology-National Science and Technology Major Project (Grant No. 2024ZD0302401 and No. 2021ZD0301500). S.L. acknowledges the China Postdoctoral Science Foundation (No. 2023M740119). G.V. acknowledges financial support from the Austrian Science Fund (FWF) through the grants P 35810-N and P 36633-N (Stand-Alone) and from the grant RYC2024-048278-I funded by MCIU/AEI/10.13039/501100011033 and FSE+. M.H. acknowledges support from the Horizon-Europe research and innovation programme under grant agreement No 101070168 (HyperSpace), and from the Austrian Federal Ministry of Education, Science and Research via the Austrian Research Promotion Agency (FFG) through the project FO999914030 (MUSIQ) and the project FO999921415 (Vanessa-QC) funded by the European Union—NextGenerationEU.

\appendix

\section{Appendix}

The appendix contains a summary of witnesses of the entanglement dimensionality vector, along with detailed proofs of the observations presented in the main text.

\subsection{Further details on known entanglement-dimensionality-vector witnesses}\label{Appendix:KnownWitnesses}

Here we present and discuss more details of known witnesses of the entanglement-dimensionality vector. 

\subsubsection{Fidelity-based witnesses}

First let us start with the fidelity witness. In the bipartite case they are among the most typical Schmidt-number witnesses used in experiments~\cite{MarioGeneration2014,Erker2017Quantifying,bavaresco2018measurements,FlammiaLiu11}.
In that case, given a target state written in its Schmidt decomposition
\be
\ket{\Psi}=\sum_i \sqrt{\lambda_i} \ket{i_a i_b} ,
\ee
where $\lambda_i$ are the squared Schmidt coefficients which are ordered non-increasingly, one can easily calculate its maximal fidelity with any state $\varrho$ with Schmidt number bounded by $r$. 
This is simply given by
\be\label{eq:maxfidLambda}
F_{\rm max}(r, \Psi) :=  \max_{\SN(\varrho) \leq r} \tr\left( \varrho \ketbra{\Psi}  \right) = \max_{\SN(\psi_r) \leq r} |\bra{\psi_r} \Psi \rangle |^2 = \sum_{i=1}^r \lambda_i ,
\ee
i.e., by the sum of the $r$ largest squared Schmidt coefficients of the target state.

Let us now elucidate how to extend this result to the multipartite case, in which many possible bipartitions have to be considered.
Let us start with the simpler question: Is the smallest Schmidt number across all possible bipartitions larger than a given $v_{\mathcal N}$? In the particular case $v_{\mathcal N} > 1$ the state is called Genuinely Multipartite Entangled. 

To obtain this fidelity bound we have to allow the possibility that the value $v_{\mathcal N}$ corresponds to the Schmidt-rank across any bipartition. 
Using \cref{eq:maxfidLambda} we know that
the fidelity between $\ket{\Psi}$ and a pure state with Schmidt-rank equal to $v_{\mathcal N}$ across bipartition $\alpha$ will be given by $\sum_{i=1}^{v_{\mathcal N}} \lambda_i^{(\alpha)}$ at most, where again $\lambda_i^{(\alpha)}$ are the non-increasingly ordered squared Schmidt coefficients of $\ket{\Psi}$ across the bipartition $\alpha$. 
Next, we want to find the maximal fidelity between the target state and any pure state that has Schmidt rank $v_{\mathcal N}$ across {\it any} of the bipartitions.
For this, we have to scan across all bipartitions, i.e., consider states $\ket{\psi}$ that have $\SN_\alpha(\psi) \leq v_{\mathcal N}$ across an arbitrary bipartition $(\alpha | \bar \alpha)$ and then take the maximal (i.e., worst case) value. This way, we get
\be\label{eq:fidmaxV}
F_{\rm max}(v_{\mathcal N} , \Psi) := \max_\alpha \max_{\SN_\alpha(\psi) \leq v_{\mathcal N}} |\bra{\psi} \Psi \rangle |^2 =\max_\alpha \left\{ \sum_{i=1}^{v_{\mathcal N}} \lambda_i^{(\alpha)} \right\} .
\ee
Now, let us observe how can we use this value to exclude that a given mixed state $\varrho$ has $\ED_{\mathcal N}(\varrho)\leq v_{\mathcal N}$, given its fidelity to $\ket{\Psi}$.
From the definition in \cref{eq:SNVdef} we see that there will exist a particular decomposition $\varrho = \sum_k p_k \ketbra{\psi_k}$ such that each $\ket{\psi_k}$ has Schmidt rank $r\leq v_{\mathcal N}$ across {\it some} bipartition. Considering such a decomposition we have:
\be
F(\varrho, \Psi) = \bra{\Psi} \varrho \ket{\Psi} = \sum_k p_k |\braket{\Psi}{\psi_k}|^2 \leq \sum_k p_k F_{\rm max}(v_{\mathcal N} , \Psi) = F_{\rm max}(v_{\mathcal N} , \Psi) .
\ee

Next, suppose that we want to find the fidelity bound with such a given target state among all possible density matrices that have 
$\ED_{\mathcal N}(\sigma)\leq v_{\mathcal N}$ and $\ED_{\mathcal N-1}(\sigma)\leq v_{\mathcal N-1}$, i.e., the two smallest Schmidt numbers are (upper bounded by) $v_{\mathcal N}$ and $v_{\mathcal N-1}$ respectively. 
Note that the bound would depend potentially on both chosen values $(v_{\mathcal N-1},v_{\mathcal N})$ independently, as one cannot find any ordering already in these two-component subvectors. 
Moreover, as the definition of the SN vector is an element-wise construction, and the optimum for each element can be drawn from different decompositions, we must consider at least two different decompositions, one providing the $\ED_{\mathcal N}(\sigma)$ and the other providing the $\ED_{\mathcal N-1}(\sigma)$. For getting the former value, as before, there must exist a decomposition with all pure states with Schmidt rank $r\leq v_{\mathcal N}$ for all bipartitions.
The latter value $\ED_{\mathcal N-1}(\sigma)\leq v_{\mathcal N-1}$ is however, only the second-to-smallest Schmidt number, and this only implies that there exists (potentially another) decomposition that has the two smallest Schmidt ranks being $(v_{\mathcal N-1}, r_\mathcal N)$ where the latter value is just $r_{\mathcal N} \leq v_{\mathcal N-1}$ (without the necessity that $r_{\mathcal N} \leq v_{\mathcal N}$). Hence, now one would have to consider pure state fidelity bounds of the form
\be
F_{\rm max}( (r_{\mathcal N-1},r_{\mathcal N}) , \Psi) := \max_{(\alpha,\beta)} \max_{\substack{\SN_\alpha(\psi) \leq r_{\mathcal{N}}, \\ \SN_\beta(\psi) \leq r_{\mathcal{N}-1}}} |\bra{\psi} \Psi \rangle |^2 \leq \max_{(\alpha,\beta)} \min \{  \sum_{i=1}^{r_{\mathcal N}} \lambda_i^{(\alpha)} , \sum_{i=1}^{r_{\mathcal N-1}} \lambda_i^{(\beta)} \} ,
\ee
with $(r_{\mathcal N-1}, r_\mathcal N)$ being all (ordered, i.e., $r_{\mathcal N-1} \geq r_{\mathcal N}$) pairs of numbers for which either $r_{\mathcal N-1} \leq v_{\mathcal N-1}$ or $r_{\mathcal N} \leq v_{\mathcal N}$.
Here, the minimization is for a fixed pair of bipartitions $(\alpha,\beta)$, in order to take the correct value of the fidelity allowed in that case, and the maximization is over the bipartitions, as we have to consider the worst-case scenario (i.e., highest-fidelity among all pairs or Schmidt ranks $(r_{\mathcal N-1}, r_\mathcal N)$ across all pairs of bipartitions).

Thus, already the pure-state optimization becomes much more involved. Moreover, in order to make use of such bounds to exclude certain SN vectors for a mixed state as we mentioned we now essentially have to consider all such bounds for all pairs of numbers $(r_{\mathcal N-1}, r_\mathcal N)$ and divide them into feasible and unfeasible. The former are those for which 
$F(\varrho, \Psi) \leq F_{\rm max}( (r_{\mathcal N-1},r_{\mathcal N}) , \Psi)$ while the latter are those for which $F(\varrho, \Psi) > F_{\rm max}( (r_{\mathcal N-1},r_{\mathcal N}) , \Psi)$.
Since now the two elements of the SN vector are associated to potentially two different decompositions, we can only conclude that 
\bea
\ED_{\mathcal N-1}(\varrho) & \geq \min_{\vec{v} \in \mathcal{V}_\text{feas}} v_{\mathcal N-1} ,\quad 
\ED_{\mathcal N}(\varrho) & \geq \min_{\vec{v} \in \mathcal{V}_\text{feas}} v_{\mathcal N} .
\eea
Similarly, we can extend this argument to more components of the SN vector and perform the analysis as described in \cref{sec:knownSNwit} and arrive at \cref{eq:feasiblesetDef}.

Let us now clarify this method with a concrete example, illustrating the most common situation, in a tripartite system with Hilbert space $\hil = \mathbb C^{3} \otimes \mathbb C^{3} \otimes \mathbb C^{3}$. Let us consider as a target the GHZ state as in \cref{eq:ghzpurestate} with $d=3$ and $N=3$, 
which has entanglement-dimensionality vector $\EDV\left(\ket{\Psi^3_{\rm GHZ}}\right)=(3,3,3)$. 
In this case we also simply have
\be
\vec \lambda^{(\alpha = 1)} = \vec \lambda^{(\alpha = 2)} = \vec \lambda^{(\alpha = 3)} = \left( \tfrac 1 3 , \tfrac 1 3 , \tfrac 1 3 \right) ,
\ee
because the state is invariant under permutation of the parties and the one-body marginals are all maximally mixed. 
Let us now try to find the maximal fidelity with $\ket{\Psi^3_{\rm GHZ}}$ among all states that have an ordered Schmidt number vector given by $\vec v = (3,3,2)$, namely calculate $F_{\rm max}(\vec v= (3,3,2) , \Psi^3_{\rm GHZ})$. For this case, we have to check all pure states that have (non-ordered) Schmidt rank vectors given by $\vec s_1 = (3,3,2)$, $\vec s_2 = (3,2,3)$ and $\vec s_3 = (2,3,3)$. For this specific target state we actually always get the same bound, which is
\be
F_{\rm max}(\vec v= (3,3,2) , \Psi^3_{\rm GHZ}) \leq \frac 2 3 .
\ee
At the same time, the symmetry of the state has also the consequence that the same bound holds for all given $\vec v$ such that their smallest element is a given $v_{\mathcal N}$ (which is equal to two in this example). Thus, with the fidelity with GHZ states we can only distinguish states with a given $v_{\mathcal N}$. In particular, for a canonical GHZ state \eqref{eq:ghzpurestate} with all particles of dimension $d$ we get the bound
\be
F_{\rm max}(v_{\mathcal N} , \Psi^d_{\rm GHZ}) \leq \frac{v_{\mathcal N}} d ,
\ee
which we are going to explain in more detail later.

\subsubsection{Entanglement-dimensionality criterion from the linear entropy vector} 

Huber et al. introduced a criterion in~\cite{HuberStructure2013,HuberPerarnaudeVicentePRA13} based on the convex-roof-extended linear entropy vector $\vec{\mathcal E}^{\rm lin}$ with elements defined in \cref{eq:Eofalphagen} and ordered non-increasingly. Let us recall their method in the following. 
Let us consider a generic $N$-particle pure state, expanded in the computational basis
\be\label{eq:purestateeta}
\ket{\Psi} = \sum_{\eta} c_{\eta}|\eta\rangle ,
\ee
where $\eta=(i_1,\dots, i_N)$ is a multi-index with $N$ entries, taking values from $0$ to $d_n-1$, where $d_n$ is the dimension of particle $n$. 
Let us now consider a bipartition $(\alpha | \bar \alpha)$ and denote by $(\eta_\alpha,\eta_\alpha^\prime)$ the pair of multi-indices that is obtained from $(\eta,\eta^\prime)$
by exchanging all indices corresponding to party $\alpha$. The linear entropy is defined as $S^{\rm lin}_\alpha(\Psi)=\sqrt{2\left[1-\tr\left(\varrho_\alpha^2\right)\right]}$ where $\varrho_\alpha$ is the reduced density matrix.

For a pure state expanded as in \cref{eq:purestateeta}, the linear entropy relative to party $\alpha$ can be expressed as
\be
(S^{\rm lin}_\alpha(\Psi))^2 = \sum_{\eta, \eta^{\prime}}\left|c_\eta c_{\eta^{\prime}}- c_{\eta_\alpha} c_{\eta_\alpha^{\prime}}\right|^2 ,
\ee
and given any subset of pairs of multi-indices $C$ can be lower bounded as
\be
S^{\rm lin}_\alpha(\Psi) \geq \frac{1}{\sqrt{|C|}} \sum_{\eta, \eta^{\prime} \in C}\left(\left|c_\eta c_{\eta^{\prime}}\right|-\left|c_{\eta_\alpha} c_{\eta_\alpha^{\prime}}\right|\right) .
\ee
Because of that, the $k$-th element of the ordered vector of linear entropies of the marginals can be lower bounded, for pure states, as
\begin{equation}
\begin{aligned}
(S^\downarrow_k)^{\rm lin}(\Psi) & \geq \frac{1}{\sqrt{|C|}} \sum_{\eta, \eta^\prime \in C}\left(\left|c_\eta c_{\eta^\prime}\right|- \min_{\mathcal R_k \subset \{1,\dots,\mathcal N\} } \sum_{m=1}^k \left|c_{\eta_{\alpha_m}} c_{\eta_{\alpha_m}^\prime}\right|\right) ,
\end{aligned}
\end{equation}
where the minimization is over all subsets of bipartitions $\mathcal R_k=\{\alpha_1,\dots ,\alpha_k \}\subset \{1,\dots,\mathcal N\}$ with cardinality equal to $k$.

This relation can be generalized to mixed states through the application of $\inf (A-B) \geq \inf A-\sup B$, which then allows to minimize over all decomposition $\mathcal{D}(\varrho)$. 
As a result one gets the following lower bound for the $k$-th component of the linear entropy vector: 
\begin{equation}
(\mathcal E^\downarrow_k)^{\rm lin}(\varrho) \geq \frac{1}{\sqrt{|C_k|}} \sum_{\eta, \eta^\prime \in C_k}\left[\left|\bra{\eta}\varrho \ket{\eta^\prime} \right|-\min_{\mathcal R_k} \sum_{m=1}^k \sqrt{\bra{\eta_{\alpha_m}}
\varrho \ket{\eta_{\alpha_m}}\bra{\eta_{\alpha_m}^{\prime}}\varrho \ket{\eta_{\alpha_m}^{\prime}}}\right] ,
\end{equation}
which depends on a chosen subset of pair of indices $C_k$, that might also be different for the different $k$.
Then the elements $v_k$ in the Schmidt number vector are bounded by using the relation
\be
(\mathcal E^\downarrow_k)^{\rm lin}(\varrho) \leq \sqrt{2\left(1-\frac{1}{v_k}\right)} ,
\ee
which must hold for all states such that 
\be
\SN^\downarrow_k(\varrho) \leq v_k .
\ee
This method thus, gives some flexibility in choosing the best set of indices $C_k$ so to optimize the detection of a given state. 
For example, for the GHZ state $\left|\Psi_{\mathrm{GHZ}}^3\right\rangle=\frac{1}{\sqrt{3}}(|000\rangle+|111\rangle+|222\rangle)$ proper choices are $C_k=\{(000,111),(000,222),(111,222)\}$, with $k=1,2,3$, which is the same set for all $k$. Still, similarly as for fidelity-based witnesses, for a general density matrix it is very demanding to find the optimal bases for evaluating this criterion, especially if no information about the state is provided in advance.

The other exemplary states that we consider are of the form \eqref{eq:432purestate}, namely
\begin{equation}
\ket{\psi_{432}}(\vec c):=c_1 \ket{000}+c_2 \ket{111}+c_3 \ket{012} +c_4 \ket{123} 
\end{equation}
mixed with white noise. For such states in our numerical calculations we consider the following choices: 
\be
\begin{aligned}
C_1&=\{(000,111),(000,123),(012,123),(000,012),(111,123),(111,012)\} , \\
C_2&=\{(000,111),(000,123),(012,123),(000,012),(111,123)\} , \\
C_3&=\{(000,111),(000,123),(012,123)\} .
\end{aligned}
\ee

\subsubsection{Detection of the single-particle rank vector from the correlation tensor norm} 

Kl\"ockl and Huber proposed a criterion for entanglement-dimensionality vector based on the $2$-norm of the correlation tensor~\cite{KlocklCharacterizing2015}, which we are going to summarize in the following.
Let us consider a $N$-qu$d$it density matrix and expand it in terms of single-particle orthonormal bases:
\begin{equation}
\varrho=
\sum_{\mu_1, \ldots, \mu_N} \av{g^{(1)}_{\mu_1} \otimes \cdots \otimes g^{(N)}_{\mu_N}} g^{(1)}_{\mu_1} \otimes \cdots \otimes g^{(N)}_{\mu_N},
\end{equation}
and consider single-particle bases $g^{(n)}_{\mu_n}$ that are composed of the identity matrix $g^{(n)}_0=\id_{d}/\sqrt{d}$ and the (normalized) generators of the $su(d)$ algebra $\{\sigma^{(n)}_1 , \dots , \sigma^{(n)}_{d^2-1} \}$.
In this generalized Bloch decomposition all relevant information about the density matrix is carried by the correlation tensor among $su(d)$ operators for all possible subsystems.
In particular, in \cite{KlocklCharacterizing2015} the authors considered the correlations of $su(d)$ observables among all particles\footnote{Note that as compared to \cite{KlocklCharacterizing2015} we now use a different normalization. In \cite{KlocklCharacterizing2015} they normalized the basis observables with $\tr(g_\mu g_\nu)=d\delta_{\mu \nu}$.}:
\be
T_{\mu_1,\dots,\mu_N} = \av{\sigma^{(1)}_{\mu_1} \otimes \cdots \otimes \sigma^{(N)}_{\mu_N}} ,
\ee
where note that the indices $\mu_n$ now run from $1$ to $d^2-1$ for each particle.
Partitioning the $N$-particle system in two parts $(\alpha | \bar \alpha)$ the marginal state with respect to party $\alpha$ composed of particles $\{n_1 , \dots , n_{|\alpha|} \} \subset \{1,\dots ,N\}$ 
is then characterized by the corresponding $|\alpha|$-body tensor:
\be
T_{\mu_1, \mu_2, \ldots, \mu_{|\alpha|}}^{(\alpha)} := \av{\sigma^{(n_1)}_{\mu_{n_1}} \otimes \cdots \otimes \sigma^{(n_{|\alpha|})}_{\mu_{n_{|\alpha|}}}} .
\ee
Let us now consider its $2$-norm, defined as
\be
\|T^{(\alpha)}\|_2:=\sqrt{\sum_{\mu_1, \mu_2, \ldots, \mu_{|\alpha|}} \left(T_{\mu_1, \mu_2, \ldots, \mu_{|\alpha|}}^{(\alpha)}\right)^2 } ,
\ee
where again the indices run from $1$ to $d^2-1$ for each particle.
We then take the $K$-particle correlation tensor norm, which is defined as
\begin{equation}
\mathcal{C}_K(\varrho):=\sum_{m=K}^N \sum_{|\alpha|=m} \left\|T^{(\alpha)}\right\|_2^2,
\end{equation}
where $m$ is the number of single parties included in the party $\alpha$. For example, the $2$-particle correlation tensor norm of a tripartite state is 
\be
\mathcal{C}_2\left(\varrho\right)=\left\|T^{(12)}\right\|_2^2+\left\|T^{(23)}\right\|_2^2+\left\|T^{(13)}\right\|_2^2+\left\|T^{(123)}\right\|_2^2 . 
\ee
Now, since $\mathcal{C}_K(\varrho)$ is convex, it is possible to find upper bounds that are valid for all pure states with a given Schmidt rank vector $\vec v$, and then those will be immediately valid for
all mixed states with entanglement-dimensionality vector given by $\vec v$. 

To find such an upper bound, the authors of \cite{KlocklCharacterizing2015} used the constraint
\be
\sum_{\mu_n} \av{\sigma^{(n)}_{\mu_n}}^2 \geqslant \frac 1 {d} \left( \frac{d}{k_n}-1\right) ,
\ee
which is valid for all pure states such that the ranks of the single-particle marginals have are $k_n$,
along with the relation
\be
\tr\left(\varrho^2\right)=\left(\sum_{m=0}^N \sum_{|\alpha|=m}\|T^{(\alpha)}\|_2^2\right)=1 , 
\ee
where now also the case zero particles subsystems $|\alpha|=0$ is formally included.

Thus, every pure state $\ket{\Psi}$ that is such that its single-particle ranks are $\left(k_1, k_2, \ldots, k_N\right)$, must satisfy the inequality
\begin{equation}\label{eq:corrtensineq_app}
\mathcal{C}_2\left(\ket{\Psi}\right) \leqslant d^N+N-1-\sum_n \frac{d}{k_n} ,
\end{equation}
which is then extended by convexity to all mixed states whose pure-state components have single-particle marginals with ranks $k_n$. We refer to it as the single-particle Schmidt numbers for simplicity. 

Note that in this case, the conclusion is not about the SN vector as defined in \cref{eq:SNVdef}. Rather,
violating \cref{eq:corrtensineq} indicates that at least one of the single-particle Schmidt numbers of $\varrho$ is greater than the corresponding rank in the vector $\left(k_1, k_2, \ldots, k_N\right)$. Different from the full entanglement-dimensionality vector, this vector only contains local ranks of single particles and moreover it refers to concrete bipartitions, without ordering the ranks non-increasingly.
In this case, in fact, the rank $k_n$ is indexed by the particle index $1\leq n \leq N$, and the assumption is that there exists a decomposition in which, for each $n$, the single-particle Schmidt number of all pure-state components does not exceed $k_n$.

\subsection{The CMC for Schmidt number and its corollary}\label{Appendix:CMCRecall}
We consider the 
(symmetric) covariance matrix, that for a generic vector of (hermitian) operators $\vec M=(M_1,\dots ,M_K)$ is defined as
\be
[\Gamma_\varrho(\vec M)]_{jk} := \tfrac 1 2 \av{M_j M_k + M_k M_j}_\varrho - \av{M_j}_\varrho \av{M_k}_\varrho . 
\ee
The covariance matrix is: (i) {\it positive} for all states $\varrho$ and all vectors of operators $\vec M$ and (ii) {\it concave} for mixing the quantum state, i.e., $\Gamma_{p\varrho_1 +(1-p)\varrho_2} \geq p \Gamma_{\varrho_1} + (1-p)\Gamma_{\varrho_2}$.

In particular, fixed a bipartition $\alpha$ we consider the covariance matrix of a couple of orthonormal bases $\obsset{g} = (\obsset{g}_\alpha, \obsset{g}_{\bar \alpha})$. Calculated on a generic mixed state $\varrho$, this assumes the block form
\begin{equation}\label{eq:blockCov}
\Cov_\varrho(\obsset{g}) := \Gamma_\varrho^{(\alpha)} = 
\left(\begin{array}{ll}
\gamma_\alpha & \C \\
\C^T & \gamma_{\bar \alpha}
\end{array}\right),
\end{equation}
in which the diagonals $\gamma_\alpha:=\Covn{\alpha}$ and $\gamma_{\bar \alpha}:=\Covn{{\bar \alpha}}$ are the covariance matrices of each party, and the off-diagonal blocks are
\be
\begin{aligned}
(\C)_{kl} &= \av{g_k^{(\alpha)} \otimes g_l^{(\bar \alpha)}}_\varrho - \av{g_k^{(\alpha)}}_\varrho \av{g_l^{(\bar \alpha)}}_\varrho ,
\end{aligned}
\ee
namely the cross-covariances between the single-party observables vectors. 
It is useful to recall that this matrix can be brought in a block singular value decomposition with a suitable local orthogonal transformation $O \Gamma_\varrho^{(\alpha)} O^T$, with $O=O_a \oplus O_b$. 
This corresponds to an orthonormal change of local bases $\obsset g \mapsto \obsset g^\prime = O \obsset g$.

Let us then consider a density matrix $\varrho$ such that its Schmidt number across bipartition labelled by $\alpha$ is $\SN_{\alpha}(\varrho) \leq r_\alpha$. 
We know~\cite{Liu2024bounding} that every such density matrix must satisfy
\be\label{eq:CMCrankr}
\Gamma_\varrho^{(\alpha)} \geq \sum_k p_k \Gamma_{\psi^{r_\alpha}_k}^{(\alpha)} ,
\ee
where $\ket{\psi^{r_\alpha}_k}$ are generic pure states with Schmidt rank smaller or equal to $r_\alpha$ across bipartition $(\alpha | \bar \alpha)$ that provide the boundary covariance matrices.
This has been proven in \cite{Liu2024bounding} by using essentially concavity of the covariance matrix, and at the same time a general form of the boundary covariance matrices for the pure Schmidt-rank-$r$ states has been provided.

Corollaries of \cref{eq:CMCrankr} were also studied in \cite{Liu2024bounding}, and especially one can write the following relation in terms of the trace norm of the blocks:
\be\label{eq:cor1app}
f_\alpha (\varrho):=\tr|\C^{(\alpha)}| - \sqrt{[1 - \tr(\varrho^2_\alpha)][1 -\tr(\varrho^2_{\bar \alpha})] } +1 \leq r_\alpha ,
\ee
which is a relation that must hold for all states that have a Schmidt number of at most $r_\alpha$ across the bipartition $(\alpha | \bar \alpha)$.
Here, $\varrho_\alpha$ is the reduced density matrix relative to party $\alpha$ and $\varrho_{\bar \alpha}$ is the reduced density matrix of its complement.
Below we repeat the idea of the proof as an illustration for the subsequent proof of our criterion in the multipartite scenario.

Consider the matrix $\DiffMat := \Gamma_{\varrho} - \sum_k p_k \Gamma^{(k)}_r$, which due to \cref{eq:CMCrankr} must be positive for all Schmidt-number-$r$ states. Here we define
\begin{equation}
\sum_k p_k \Gamma^{(k)}_r:=
\left(\begin{array}{cc}
\kappa_\alpha & X_{\psi_r} \\
X_{\psi_r}^T & \kappa_{\bar{\alpha}}
\end{array}\right).
\end{equation}
Positivity of a block matrix implies the inequality (see e.g., \cite{HornJohnson})
\be\label{eq:hyperb}
\tr |\DiffMat_\alpha | \cdot \tr | \DiffMat_{\bar \alpha} | \geq \tr | (\DiffMat_X^T\DiffMat_X)^{1/2} |^2 ,
\ee
where we have labelled the blocks $\DiffMat_{\alpha,\bar \alpha,X}$ in analogy with a generic covariance matrix. The above inequality is equivalent to the following family of inequalities
\be\label{eq:ineqwithtini}
\tr(\DiffMat_\alpha) + 4 t^2 \tr(\DiffMat_{\bar \alpha}) \geq 4 |t| \tr|\DiffMat_X| \geq 4 |t| (\tr|X_\varrho| - \tr|X_{\psi_{r}}|),
\ee
where $t$ is a real parameter. Here, in the last inequality we substituted the expression $\DiffMat_X=X_\varrho - X_{\psi_{r}}$ and used the triangle inequality. 

Since $\DiffMat_\alpha$ and $\DiffMat_{\bar \alpha}$ are positive (being principal minors of $\DiffMat$) we have 
\be\label{eq:diffmatalphaapp}
\tr(\DiffMat_\alpha)= \tr(\gamma_{\alpha}) - \tr(\kappa_{\alpha}) = 1-\tr(\varrho^2_{\alpha})-E_L(\psi_{r})
\ee
and analogously for $\tr(\DiffMat_{\bar \alpha})$ where $E_L(\psi)=1-\sum_k (\lambda_\psi)^2_k$ is the linear entanglement entropy of a pure bipartite state with squared Schmidt coefficients given by $(\lambda_\psi)_k$. To derive \cref{eq:diffmatalphaapp} we used that 
\be
\tr(\gamma_\varrho) = d - \tr(\varrho^2) 
\ee
holds for a generic single qu$d$it covariance matrix and we consider the generic (optimal) pure Schmidt rank-$r$ state $\ket{\psi_{r}}$.

The rest of the proof consists basically in exploiting the bound 
\be
\tr|X_{\psi_r}|\leq r- 1 + E_L(\ket{\psi_r}) ,
\ee
which was proven in \cite{Liu2024bounding}. Substituting all the relations above into \cref{eq:ineqwithtini} we obtain that the following inequality
\be\label{eq:condswitht}
1-\tr(\varrho^2_\alpha)-E_L(\ket{\psi_{r}}) + 4 t^2 (1-\tr(\varrho^2_{\bar \alpha})-E_L(\ket{\psi_r}))\geq 4 |t| (\tr|X_\varrho| - r+ 1 - E_L(\ket{\psi_r})) , 
\ee
holds for all values of $t$. 
Minimizing the left-hand side over $t$ we get that the minimum is achieved for  $2|t|=(\tr|X_\varrho|-r+1)/(1-\tr(\varrho_{\bar \alpha}^2))$ and results in \cref{eq:cor1app}.

Afterwards, we observe that given the inequality \eqref{eq:cor1app} one can find weaker conditions. For example, by using the inequality between arithmetic and geometric mean, we can find the bound 
\be
f_\alpha (\varrho) \geq \tr|\C^{(\alpha)}| + \frac{\tr(\varrho^2_\alpha) + \tr(\varrho^2_{\bar \alpha})} 2 \geq \tr(\C^{(\alpha)}) + \frac{\tr(\varrho^2_\alpha) + \tr(\varrho^2_{\bar \alpha})} 2 ,
\ee
($X_{\varrho}^{(\alpha)}$ can be extended to a square matrix by augmenting the system dimension) which would lead to the criterion:
\be
\tr(\C^{(\alpha)}) + \frac{\tr(\varrho^2_\alpha) + \tr(\varrho^2_{\bar \alpha})} 2 \leq r_\alpha .
\ee
Note that this criterion can be also directly obtained from \cref{eq:condswitht} for $t=1/2$.
Interestingly, an even weaker criterion is related to the fidelity with respect to states that are maximally entangled across the bipartition $\alpha$ and reads
\be\label{eq:weakerthancor1app}
\sum_{K=1}^{d_\alpha^2} \av{g^{(\alpha)}_{K}\otimes g^{({\bar \alpha})}_{K}}_\varrho \leq r_\alpha ,
\ee
where $d_\alpha$ is a shorthand notation for $\min \{d_\alpha , d_{\bar \alpha}\}$, i.e., the smaller of the two dimensions between the parties and $\{ g_\mu^{(\alpha)}\}$, $\{ g_\mu^{({\bar \alpha})}\}$ are basis of observables for the two parties. 
To prove \cref{eq:weakerthancor1app} we simply observe that 
\be
\begin{aligned}\label{eq:weakerthancorrderivation}
&\tr(\varrho^2_\alpha)+\tr(\varrho^2_{\bar \alpha}) + 2\tr|\C^{(\alpha)}| \geq \sum_{K=1}^{d^2_\alpha} \left( \av{g^{(\alpha)}_{K}}^2+ \av{g^{({\bar \alpha})}_{K}}^2- 2 \av{g^{(\alpha)}_{K}}\av{g^{({\bar \alpha})}_{K}} + 2 \av{g^{(\alpha)}_{K}\otimes g^{({\bar \alpha})}_{K}} \right) \\
&= \sum_{K=1}^{d^2_\alpha} \left[ \left( \av{g^{(\alpha)}_{K}} - \av{g^{({\bar \alpha})}_{K}} \right)^2 + 2 \av{g^{(\alpha)}_{K}\otimes g^{({\bar \alpha})}_{K}} \right] 
\geq 2 \sum_{K} \av{g^{(\alpha)}_{K}\otimes g^{({\bar \alpha})}_{K}} ,
\end{aligned}
\ee
where $\{ g^{(\alpha)}_{K}\}$ and $\{ g^{(\bar \alpha)}_{K}\}$ are optimally chosen as the bases that bring $\C^{(\alpha)}$ in its singular-value decomposition.
Note also that we can derive a bound similar to the above by considering any other two bases $\{ \tilde g^{(\alpha)}_{K}\}$ and $\{ \tilde g^{(\bar \alpha)}_{K}\}$ and also
by discarding some of the indices $K$. This is because each of the terms inside the sum in \cref{eq:weakerthancorrderivation} comes from the singular value decomposition of $\C^{(\alpha)}$ and is thus positive.

When there is only one bipartition, i.e., in the bipartite case, the criterion in \cref{eq:weakerthancor1app} is related to the fidelity with respect to a (optimal) maximally entangled state
\be
\ket{\Psi^d}\bra{\Psi^d} = \frac{1}{d}
\sum_K g_{K}\otimes g_{K} ,
\ee
that is constructed from the optimally chosen bases $\{ g^{(\alpha)}_{K}\}$ and $\{ g^{(\bar \alpha)}_{K}\}$~\cite{Liu2024bounding}.

\subsection{Proof of \cref{eq:obs2sys}}\label{Appendix:ProofObs2sys}

Let us consider a mixed state $\varrho$ such that 
we can find a decomposition of the form 
\be\label{eq:decompoSNVapp}
 \varrho = \sum_k p_k \varrho_{\vec s_k} ,
\ee
where $p_k$ are probabilities and $\varrho_{\vec s_k}$ are pure states with a given (unordered) Schmidt number vector $\vec s_k$ that is such that $(s^\downarrow_k)_j \leq v_j$ when the components of $\vec s_k$ are ordered non-increasingly. 
Using the concavity of the covariance matrix, we can derive a matrix inequality for the covariance matrix of $\varrho$ relative to the bipartition $\alpha$, 
which we write as
\be
\Gamma_\varrho^{(\alpha)} \geq \sum_k p_k \Gamma^{(\alpha)}_{\vec s_k} := \Gamma^{(\alpha)}_{\vec v} ,
\ee
where the bound on the right-hand side contains a mixture of covariance matrices of pure states with different Schmidt numbers $s_{k ,\alpha}$. 

For these pure state boundary covariance matrices, we can use the relation
\be
\tr | \Cp | \leq s - 1 + E_L(\psi) ,
\ee
and summing up all the terms in the decomposition in \cref{eq:decompoSNVapp} we get that
\be
\tr|X_{\vec v}^{(\alpha)}| \leq \sum_k p_k \left( s_{k ,\alpha} - 1 + E_L(\psi_{k,\alpha}) \right) ,
\ee
where $X_{\vec v}^{(\alpha)}$ denotes the off-diagonal block of the boundary covariance matrix $\Gamma^{(\alpha)}_{\vec v}$.

The diagonal blocks of $\Gamma^{(\alpha)}_{\vec v}$ are given by
\be
\begin{aligned}
    \kappa^{(\alpha)}_{\vec v} &= \sum_k p_k \kappa^{(\alpha)}_{\vec s_k} , \\
        \kappa^{(\bar \alpha)}_{\vec v} &= \sum_k p_k \kappa^{(\bar \alpha)}_{\vec s_k} ,
\end{aligned}
\ee
where $\kappa^{(\alpha)}_{\vec s_k}$ and $\kappa^{(\bar \alpha)}_{\vec s_k}$ are local covariance matrices of pure states with Schmidt number vector given by $\vec s_k$. 
Thus, we can again bound their traces by using the relation
\be
\begin{aligned}
\tr(\kappa^{(\alpha)}_{\vec s_k}) &= d_\alpha - 1 + E_L(\psi_{k,\alpha}) , \\
\tr(\kappa^{(\bar \alpha)}_{\vec s_k}) &= d_{\bar \alpha} - 1 + E_L(\psi_{k,\alpha}) ,
\end{aligned}
\ee
where $E_L(\psi_{k,\alpha})$ is the linear entanglement entropy of the pure boundary state $\ket{\psi_{k,\alpha}}$, that has Schmidt number vector $\vec s_k$. 

Thus, the trace norms of the diagonal blocks of $\Gamma^{(\alpha)}_{\vec v}$ (which are positive) are given by
\be
\begin{aligned}
\tr | \kappa^{(\alpha)}_{\vec v} | &=  \tr( \kappa^{(\alpha)}_{\vec v} ) = \sum_k p_k \tr(\kappa^{(\alpha)}_{\vec s_k})= \sum_k p_k d_\alpha - 1 + \sum_k p_k E_L(\psi_{k,\alpha}), \\
\tr | \kappa^{(\bar \alpha)}_{\vec v} | &=  \tr( \kappa^{(\bar \alpha)}_{\vec v} ) = \sum_k p_k \tr(\kappa^{(\bar \alpha)}_{\vec s_k}) = \sum_k p_k d_{\bar \alpha} - 1 + \sum_k p_k E_L(\psi_{k,\alpha}) .
\end{aligned}
\ee
Using these bounds and following the steps of the proof of \cref{eq:cor1app} we get to the bound
\be
\tr| \C^{(\alpha)} | - \sqrt{\left(1 - \tr(\varrho_{\alpha}^2) \right) \left(1 - \tr(\varrho_{\bar \alpha}^2) \right)} \leq R_\alpha - 1 ,
\ee
where $R_\alpha = \sum_k p_k s_{k ,\alpha}$. Considering all the inequalities of this form for all $\alpha$ we also get the constraints on the full vector $\vec R$ to be 
$\vec R \prec \vec v$, i.e., 
\be
\sum_{l=1}^K  R^\downarrow_l \leq \sum_{l=1}^K \sum_k p_k s^\downarrow_{k , l} \leq \sum_{l=1}^K  v_l .
\ee
This is due to the fact that the vector $\vec R^\downarrow = (\sum_k p_k \vec s_{k})^\downarrow = M \vec R$ where $M$ is a doubly stochastic matrix and the elements of the vector $\vec s_{k}^\downarrow$ are
upper bounded by the elements of $\vec v$.
\qed

\subsection{Proof of \cref{eq:weakerthancor1}}\label{Appendix:ProofWeakerthancor1}

Here we derive a condition analogous to \cref{eq:weakerthancor1app} for the multipartite case. This case is more complex because the various bipartitions have different bases with different dimensions. In such a case, one way to relate the expression in \cref{eq:weakerthancor1app} to fidelities is as follows. Let us consider the bases of the parties $\alpha$ and $\bar \alpha$ constructed from the single-particle bases:
\be
\{g^{(\alpha)}_{K}\} = \{ g_\mu^{(n)} \otimes  g_\nu^{(m)} \otimes \dots \}_{n,m\dots \in \alpha} ,
\ee
and similarly for $\{g^{(\bar \alpha)}_{K}\}$. Let us also consider only the indices $K=(\mu, \mu , \mu \dots)$ with $1\leq \mu \leq d^2$ and $d=\min_n d_n$ being the minimal dimension among the particles. 

Let us now consider the criterion in \cref{eq:weakerthancor1app}, now for pure states that have Schmidt rank at most $r_\alpha$ across bipartition $\alpha$. By considering the above bases with the restricted set of indices, \cref{eq:weakerthancor1app} can be rewritten as
\be
\sum_{\mu=1}^{d^2} \av{g^{(\alpha)}_{K}\otimes g^{({\bar \alpha})}_{K}}_\varrho = \sum_{\mu=1}^{d^2} \av{g_\mu^{(1)} \otimes \dots \otimes g_\mu^{(N)}}_\varrho \leq r_\alpha ,
\ee
and in this way the left-hand side is the same for all bipartitions. 
This is important because in this way we get a condition that can be minimized over all $\alpha$ and leads to 
\be\label{eq:witnfidboundapp}
\sum_{\mu=1}^{d^2} \av{g_\mu^{(1)} \otimes \dots \otimes g_\mu^{(N)}} \leq \min_\alpha r_\alpha := v_{\mathcal N} ,
\ee
which is thus a criterion that must be valid for all pure states such that their minimal entanglement dimensionality is smaller than or equal to $v_{\mathcal N}$.
The other advantage now is that the left-hand side is linear under mixing the quantum state. Thus the same criterion remains valid for all density matrices of the form 
\be
\varrho = \sum_k p_k \varrho_{\vec s_k} ,
\ee
where the $\varrho_{\vec s_k}$ are pure states such that their minimal entanglement dimensionality across all bipartitions is upper bounded by $v_{\mathcal N}$.

Afterwards, we can also observe that such a criterion relates to fidelities with $1$-uniform states.
To observe this let us consider the following state:
\be
\ketbra{\Psi^d_{\rm 1-uni}} = \frac 1 d \sum_{\mu=1}^{d^2} g_\mu^{(1)} \otimes \dots \otimes g_\mu^{(N)} .
\ee
It is easy to see that the fidelity between a density matrix $\varrho$ and $\ket{\Psi^d_{\rm 1-uni}}$ is given by
\be\label{eq:fidvarrhopsiNd}
\tr\left( \varrho \ketbra{\Psi^d_{\rm 1-uni}}\right) = \frac 1 d \sum_{\mu=1}^{d^2} \av{g_\mu^{(1)} \otimes \dots \otimes g_\mu^{(N)}}_\varrho .
\ee
Now, let us also derive the bound on such a fidelity for states with a given entanglement-dimensionality vector $\vec v$.
As we explained earlier, we have to maximize the overlap between $\ket{\Psi^d_{\rm 1-uni}}$ and any pure state $\ket{\Phi_{\vec r}}$ with ordered Schmidt rank vector given by $\vec v$
and this is obtained by considering the Schmidt decompositions of $\ket{\Psi^d_{\rm 1-uni}}$ for all bipartitions.

We consider here for simplicity the case in which the state is invariant under permutation of the particles. This is obtained when the matrices $g^{(n)}_\mu$ are equal for all the parties. Thus, to understand its Schmidt rank across different bipartitions what matters is just the number of particles in each given party. 
For example, let us consider the bipartition $\alpha = (1|2\dots N)$. The vector of squared Schmidt coefficients of $\ket{\Psi^d_{\rm 1-uni}}$ across this bipartition is given by:
\be
\vec \lambda^{(\alpha = 1)} (\Psi^d_{\rm 1-uni}) = \left( \tfrac 1 d , \dots \tfrac 1 d , 0 , \dots , 0 \right) ,
\ee
where the number of coefficients equal to $1/d$ is $d$.
The same vector of squared Schmidt coefficients we would get for every bipartition that is of the form $(1|N-1)$, i.e., one particle is on party $a$ and $N-1$ particles are on party $b$.
Actually the same vector of squared Schmidt coefficients arises for every bipartition, i.e., there are always just $d$ nonzero values of $\lambda^{(\alpha)}_k$, which are all equal to $1/d$.
Thus, the Schmidt-rank vector of such a state is given by
\be
\vec v(\Psi^d_{\rm 1-uni}) = (d,\dots , d) .
\ee
Now, let us look for the maximal fidelity of such a state with any pure $N$-qu$d$it state that is such that its Schmidt rank across the bipartition labelled by $\alpha$ (e.g., $\alpha = (1|2\dots N)$) is equal to $r_\alpha$.
Again, this is given by 
\be
\sum_{k=1}^{r_\alpha} \lambda^{(\alpha)}_k = \frac{r_\alpha} d ,
\ee
and the same expression is obtained for all bipartitions.
Thus, we have that the maximal overlap between $\ket{\Psi^d_{\rm 1-uni}}$ and any pure state $\ket{\Phi_{\vec r}}$ with Schmidt rank vector given by $\vec r = (r_1,\dots , r_{\mathcal N})$ is given by
\be
\left|\braket{\Phi_{\vec r}}{\Psi^d_{\rm 1-uni}}\right|^2 \leq \min_\alpha \sum_{k=1}^{r_\alpha} \lambda^{(\alpha)}_k = \min_\alpha \frac{r_\alpha} d ,
\ee
which then leads to the fidelity bound:
\be
\tr\left( \varrho \ketbra{\Psi^d_{\rm 1-uni}} \right) = \frac 1 d \sum_{\mu=1}^{d^2} \av{g_\mu^{\otimes N}}_\varrho \leq \min_\alpha \frac{r_\alpha} d .
\ee
Here in the equality we used the expression in \cref{eq:fidvarrhopsiNd}. We have thus observed that \cref{eq:witnfidboundapp} is equivalent to a fidelity bound with respect to a $1$-uniform state, of which the GHZ state is an example.
Note also that, for a given $\varrho$ one can consider many possible bases $g_\mu^{(n)}$ and thus many possible corresponding fidelities to target $1$-uniform states. The optimal case is given by a basis of this type that is also such that the matrix $\C^{(\alpha)}$ with $\alpha = (n| 1\dots n-1 n+1 \dots N)$ is in its singular value decomposition. Hence, this can be obtained by performing a singular value decomposition of the different $\C^{(\alpha)}$ for all the bipartitions of the form $(n| 1\dots n-1 n+1 \dots N)$.

\subsection{Further details on the states detected by our method}\label{Appendix:SampleStates}

Here we provide more details on how we generated the random samples for our numerical applications of our method. First, we note that a commonly chosen random state sampling method is given by $\varrho=U \Lambda U^{\dagger}$, where $U$ is a $D\times D$ Haar-random unitary and $\Lambda$ is a $D\times D$ real diagonal nonnegative matrix with elements that follow Lebesgue measure~\cite{VolumeKarol1998}. In our case, we consider in particular the case $d=3$. When considering a $3\times 3\times 3$ state, almost all samples closely resemble a maximally mixed state, with an entanglement-dimensionality vector $(1,1,1)$. The probability of selecting a $(3,3,3)$ vector is extremely low, making it inefficient for evaluating the performance of the different criteria. This problem also applies to the Hilbert-Schmidt metric and the Bures metric, with detailed construction methods available in the supplementary material of~\cite{WeilenmannPRL2020BeyondFidelity}. Thus, for applying our method to purely randomly chosen states we slightly modify $\Lambda$. We fix the first diagonal element $\Lambda_1$, which can be understood as the proportion of the dominant pure state component. Thus, we consider the eigenvalue matrix of $\varrho$ to be of the form:
\begin{equation}\label{eq:RandomDiagonalMatrixWithP}
\Lambda=\left(\begin{array}{cccc}
\Lambda_1 & 0 & \cdots & 0 \\
0 & \ast & \cdots & 0 \\
\vdots & \vdots & \ddots & \vdots \\
0 & 0 & 0 & \ast
\end{array}\right),
\end{equation}
with the constraint $\tr(\Lambda)=\tr(\varrho)=1$ in order to be a valid quantum state.
The value of $\Lambda_1$ is chosen randomly following a uniform distribution over the interval $[0,1]$. The remaining elements instead follow the Lebesgue measure. This approach allows us to create states with different purity more efficiently.

Whenever instead, we do not need to generate necessarily multipartite high-dimensional entangled states, we can also use the more standard method of generating all elements of $\Lambda$ at random following the Lebesgue measure. We apply this more standard method for example for considering a mixture of the canonical Greenberger-Horne-Zeilinger state with some noise, which does not necessarily need to be highly entangled, but rather close to separable.

As an additional important case-study for the application of our method, we consider states obtained as mixtures of the canonical GHZ-state with randomly generated density matrices, interpreted as random noise. 
This is motivated by the GHZ-state being the most common example of high-dimensional multipartite entangled states, and also a very important target for experimental implementation~\cite{erhard2018experimental,Bouwmeester1999,pan2000experimental,Zhaoetal2003,lu2007experimental,gao2010experimental,CerveraExperimental2022}. 
This is especially important because of its usefulness for several tasks in which distribution of entanglement is required among many parties, for example error-correction~\cite{ScottPRA2004,ArnaudCerf2013,GoyenechePRA2014} and quantum communication~\cite{malik2016multi}. 
In fact, known witnesses for genuine multipartite entanglement are often tested on these states~\cite{tothguhnePRL05,Guhne_2010,Huber_2010}.

In particular, we consider the $3\times 3 \times 3$-dimensional canonical GHZ state $\varrho_{\mathrm{GHZ}}^{\mathrm{d}=3}$ mixed with a random density matrix $\varrho_{\text{random}}$: 
\begin{equation}\label{eq:GHZwithrandomnoise}
\varrho=p \varrho_{\mathrm{GHZ}}^{\mathrm{d}=3}+(1-p) \varrho_{\text{random}} ,
\end{equation}
where the probability $p$ is also random, uniformly distributed in $[0,1]$.
The random state $\varrho_{\text{random}}$ is in this case constructed with the standard method described in~\cite{VolumeKarol1998} and summarized in the paragraph above. The results, based on 10000 sampled states, are shown in \cref{table:2}. As one might intuitively expect, a higher mixing probability $p$ leads to better preservation of $\varrho_{\mathrm{GHZ}}^{\mathrm{d}=3}$, resulting in a higher hierarchy of the Schmidt number vector. In this case our method also improves over the canonical GHZ fidelity-witness, which has been perfectly tailored to detect the noiseless target state.

\bibliographystyle{quantum}
\bibliography{references.bib}

\end{document}